\documentclass[pra,floatfix,twocolumn,superscriptaddress,showpacs,preprintnumbers,longbibliography,nofootinbib]{revtex4-1}
\usepackage{dcolumn}    
\usepackage{bm} 
\usepackage{graphicx}
\usepackage{amsmath}    
\usepackage{enumitem}

\usepackage{latexsym}
\usepackage{amsfonts}   
\usepackage{amssymb}
\usepackage{array}      
\usepackage{epsfig}
\usepackage{braket}
\usepackage{color}
\usepackage[colorlinks=true,linkcolor=blue,urlcolor=blue,citecolor=blue,pdfusetitle]{hyperref}
\usepackage{hyperref}
\usepackage[T1]{fontenc}
\usepackage[sc]{mathpazo}
\newcommand{\ra}{\rangle}
\newcommand{\la}{\langle}

\newcommand{\Tr}{\textrm{Tr}}

\newcommand{\HH}{\hat{\mathcal H}}
\newcommand{\U}{\hat{\mathcal{U}}}
\newcommand{\V}{\hat{V}}
\newcommand{\A}{\hat{\mathcal A}}
\newcommand{\Ap}{\hat{\mathcal A'}}

\newcommand{\de}{\delta}

\newcommand{\eq}{Eq.~}
\newcommand{\eqs}{Eqs.~}
\newcommand{\fig}{Fig.~}
\newcommand{\figs}{Figs.~}
\newcommand{\cf} {cf.~}

\newcommand{\eg} {e.g.~}

\newcommand{\rref} {Ref.~}
\newcommand{\rrefs} {Refs.~}

\DeclareRobustCommand\openzero{\leavevmode\hbox{0\kern-.55em0}}

\newcommand{\D}{\Delta t}

\begin{document}
\title{Collisional picture of quantum optics with giant emitters}
       
\author{Dario~Cilluffo}
\affiliation{Universit\`a  degli Studi di Palermo, Dipartimento di Fisica e Chimica -- Emilio Segr\`e, via Archirafi 36, I-90123 Palermo, Italy}
\affiliation{NEST, Istituto Nanoscienze-CNR, Piazza S. Silvestro 12, 56127 Pisa, Italy}
\author{Angelo~Carollo}
\affiliation{Universit\`a  degli Studi di Palermo, Dipartimento di Fisica e Chimica -- Emilio Segr\`e, via Archirafi 36, I-90123 Palermo, Italy}
\affiliation{Radiophysics Department, National Research Lobachevsky State University of Nizhni Novgorod, 23 Gagarin Avenue, Nizhni Novgorod 603950, Russia}
\author{Salvatore~Lorenzo}
\affiliation{Universit\`a  degli Studi di Palermo, Dipartimento di Fisica e Chimica -- Emilio Segr\`e, via Archirafi 36, I-90123 Palermo, Italy}
\author{Jonathan~A.~Gross}
\affiliation{Institut quantique, Universit\'e de Sherbrooke, Sherbrooke QC J1K 2R1, Canada}
\author{G.~Massimo~Palma}
\affiliation{Universit\`a  degli Studi di Palermo, Dipartimento di Fisica e Chimica -- Emilio Segr\`e, via Archirafi 36, I-90123 Palermo, Italy}
\affiliation{NEST, Istituto Nanoscienze-CNR, Piazza S. Silvestro 12, 56127 Pisa, Italy}
\author{Francesco~Ciccarello}
\affiliation{Universit\`a  degli Studi di Palermo, Dipartimento di Fisica e Chimica -- Emilio Segr\`e, via Archirafi 36, I-90123 Palermo, Italy}
\affiliation{NEST, Istituto Nanoscienze-CNR, Piazza S. Silvestro 12, 56127 Pisa, Italy}

\begin{abstract}
The effective description of the weak interaction between an emitter and a bosonic field as a sequence of two-body collisions provides a simple intuitive picture compared to traditional quantum optics methods as well as an effective calculation tool of the joint emitter-field dynamics. Here, this collisional approach is extended to many emitters (atoms or resonators), each generally interacting with the field at many coupling points (``giant'' emitter). In the regime of negligible delays, the unitary describing each collision in particular features a contribution of a chiral origin resulting in an effective Hamiltonian.
The picture is applied to derive a Lindblad master equation (ME) of a set of giant atoms coupled to a (generally chiral) waveguide field in an arbitrary white-noise Gaussian state, which condenses into a single equation and extends a variety of quantum optics and waveguide-QED MEs. The effective Hamiltonian and jump operators corresponding to a selected photodetection scheme are also worked out.
\end{abstract}
\date{\today}
\maketitle

\section{Introduction}

A major focus of quantum optics is the interaction of quantum emitters, such as (artifical) atoms or resonators, with a field modeled as a continuum of bosonic modes. Accordingly, describing the dynamics generally requires to keep track of all the field modes, a task which at times can be circumvented when the focus is the open dynamics of the emitters, provided that a master equation is preliminarily derived and ensured to be completely positive. This tool is yet insufficient and must be complemented with appropriate field equations whenever one is interested in the dynamics of photons.

\begin{figure}
\includegraphics[width=0.47\textwidth]{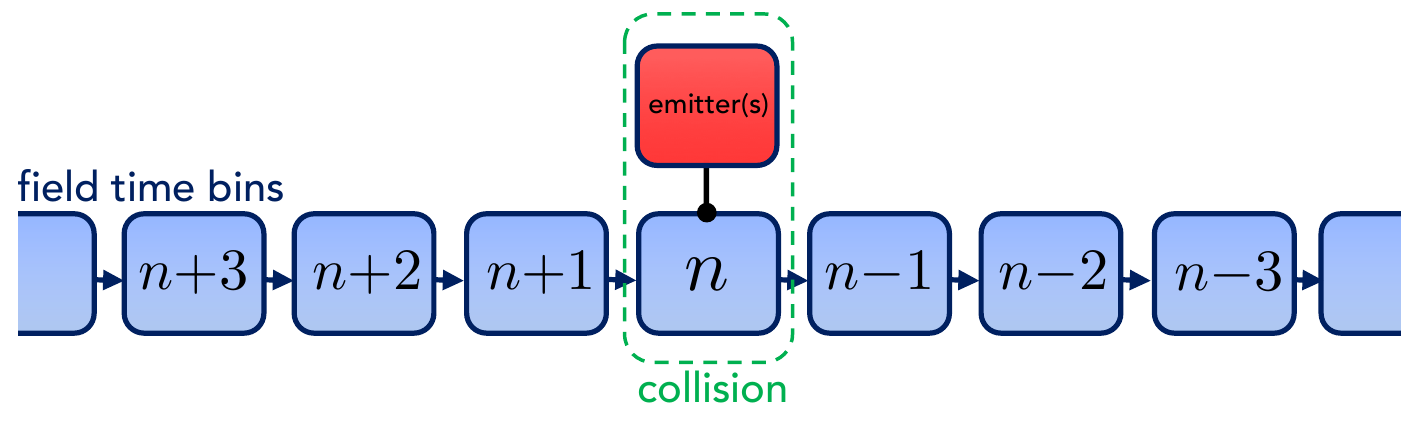}
\caption{Basic collision-model description of the emitter-field dynamics. The field is decomposed into non-interacting time bins traveling at constant speed. One at a time, these undergo a short two-body interaction with the emitter (collision). In the regime of negligible time delays, a similar conveyor-belt picture holds for many emitters each of which can be giant (i.e., interacting with the field at many coupling points).}\label{sketch}
\end{figure}

A somewhat unconventional method to tackle quantum optics problems is a collision-model description, an approach adopted in a growing number of works \cite{pichlerPhotonic2016, GrimsmoPRL15, WhalenQST17, ciccarelloCollision2017, heraus2018, FischerQ18, FischerJPC18, grossQubit2018, WhalenPRA2019, DabrowskaPRA2017, Bouten2019, DabrowskaJPA2019, DabrowskaJOSA2020, JacobsPRA20,JacobsPRL20}.
Much like in standard theories of photon counting statistics, the basic idea (see \fig\ref{sketch}) is decomposing the field into discrete time bins (each with an associated bosonic mode). In the interaction picture, time bins travel at constant speed so as to ``collide''  one at a time with the quantum emitter (conveyor-belt-like dynamics). This reduces the complex emitter-field interaction to a sequence of elementary two-body collisions, each involving a different time bin: a dynamics known in some literature as ``collision model'' (CM) or ``repeated interactions model''. CMs are being routinely used in various areas such as weak continuous measurements \cite{cavesQuantummechanical1987,brunSimple2002}, non-Markovian quantum dynamics \cite{giovannettiMaster2012a,rybarSimulation2012,ciccarelloCollisionmodelbased2013,bernardesEnvironmental2014,mccloskeyNonMarkovianity2014,JinPRA15,kretschmerCollision2016,lorenzoClass2016,lorenzoComposite2017,filippovDivisibility2017,StevePRA18}, quantum thermodynamics \cite{scaraniThermalizing2002,karevskiQuantum2009,uzdinMultilevel2014,lorenzoLandauer2015,EspositoPRX,chiaraReconciliation2018} and even quantum gravity \cite{kafriNoise2013,altamiranoUnitarity2017}. 

The CM-based description has a number of interesting features such as:
\begin{enumerate}
	\item A simple and intuitive picture of the joint dynamics, helpful to get insight into the problem at hand.
	\item A direct, Born-Markov-approximation-free, derivation of Lindblad MEs guaranteed to be completely positive. 
	\item The time-bin evolution is easily worked out, thus enabling to keep track of a relevant part of the field dynamics.
	\item CMs are the natural microscopic framework to describe continuous weak measurements, which can be applied to photon detection \cite{cavesQuantummechanical1987,brunSimple2002,altamiranoUnitarity2017,grossQubit2018}.
	\item When formulated as a CM, the dynamics turns into an equivalent quantum circuit, allowing in particular for Matrix Product States simulations \cite{schollwock2011density,pichlerPhotonic2016,grossQubit2018,Bouten2019,mahmoodian2019dynamics,Grimsmo20}.
\end{enumerate}

In the framework of quantum optics, so far only CMs for pointlike quantum emitters were fully developed (only one coupling point). While CMs featuring two coupling points were considered in the regime of long time delays \cite{pichlerPhotonic2016,GrimsmoPRL15,heraus2018}, a comprehensive formulation of the negligible-delay regime (occurring in most experiments) is still missing.

In this work, we present a general theory of the CM-based description of quantum optics in the case of many emitters. We allow each of these to generally couple to the field at many coupling points so as to encompass systems such as the so called ``giant'' atoms \cite{GuarXiv17,Kockum5years}, which can now be experimentally implemented and operated \cite{GustafssonSci14,OliverGiant2019}, or bosonic oscillators/atomic ensembles coupled to 1D fields in looped geometries \cite{AspelmeyerOptomechRMP14,HammererRMP10} as explicitly discussed in \rref \cite{Hammerer2019}. The framework is first formulated by considering a unidirectional field (just like in standard input-output formalism \cite{GardinerBook2004}) and then extended to a bidirectional field.
While both the regimes of negligible and long time delays are discussed, our main focus is the former. In which case, it will be proven that each collision can be {effectively} represented as a collective coupling of all the emitters with one field time bin plus an internal coherent dipole-dipole interaction between the emitters described by a Hamiltonian originating from the intrinsic system's chirality (in the conveyor-belt picture of \fig1 time bins travel from left to right).

While the presented collisional framework has many potential uses, here we apply it to derive the Lindblad master equation of a set of giant emitters coupled to a, generally chiral, one-dimensional waveguide when the field starts in an arbitrary Gaussian state. This condenses in a single equation and extends a variety of master equations used in waveguide QED \cite{ShiPRA15,RoyRMP17,LiaoPhyScr16,GuarXiv17}, as will be illustrated in detail. Moreover, we show that the recently discovered possibility to realize decoherence-free Hamiltonians with giant emitters \cite{KockumPRL2018,Hammerer2019} is naturally predicted in the collisional picture, without the need to resort to the master equation, thus highlighting its independence of the field state. Additionally, for an arbitrary photodetection scheme, we calculate the Kraus operators corresponding to a measurement outcome and use these to derive the effective Hamiltonian and jump operators generating the quantum trajectories.

The present paper  in fact comprises two parts. The first of which presents the general emitters-field microscopic model (Section \ref{model}), outlines the main collision model features without proof, the aforementioned general master equation and the description of photodetection and related quantum trajectories (Section \ref{summary}). Special cases of the master equation are illustrated in a separate section (Section \ref{examples}), which ends with a discussion of decoherence-free Hamiltonians (Section \ref{section-DF}).

The second (more technical) part derives in detail the collision model for a unidirectional field (Section \ref{derivation}), works out the ensuing master equation (when existing) in the negligible-delays regime (Section \ref{section-ME}), extend these tasks to a bidirectional field (Sections \ref{CM-bi} and \ref{section-ME-bi}) and finally addresses in detail photodetection and quantum trajectories (Section \ref{sec-jump}).

\section{Microscopic model} \label{model}

The general emitters-field microscopic model we consider is essentially the same as that underpinning the standard input-output formalism of quantum optics \cite{GardinerBook2004} and related theories such as SLH \cite{CombesAdvPhyX17}.

\begin{figure}
	\includegraphics[width=0.42\textwidth]{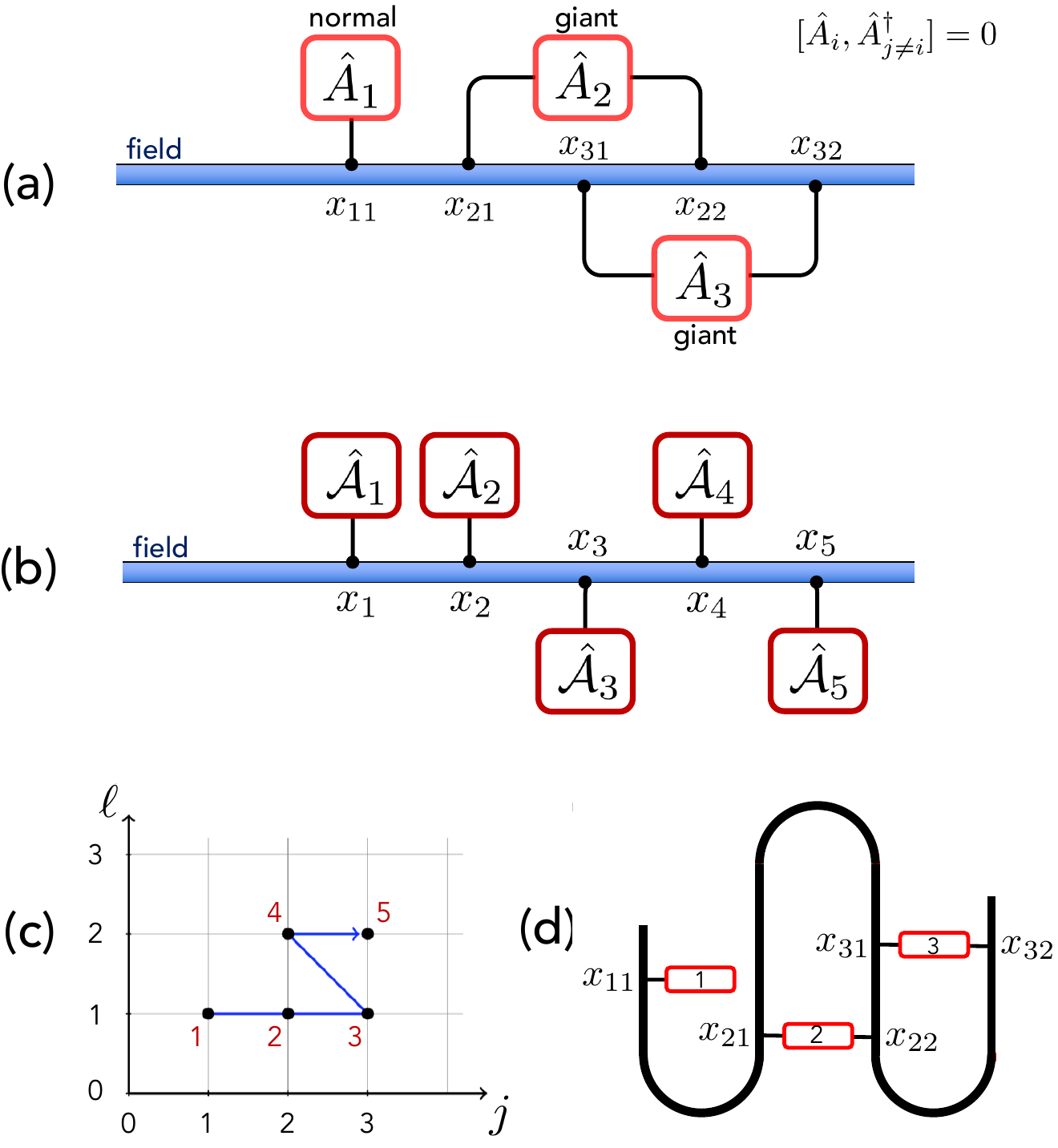}
	\caption{A set of (generally giant) emitters coupled to a unidirectional field. (a): A normal emitter (such as 1) interacts with the field at a single coupling point, while a giant emitter has two or more coupling points (like emitters 2 and 3 here). (b): Instead of a double index as in (a), we can use a single index $\nu$ to label coupling points from left to right, defining for each a ladder operator $\hat A_\nu$ incorporating the coordinate-dependent phase factor (e.g., $\hat{{\cal A}}_4=e^{-i k_0 x_{22}}\hat A_2$). Thus, formally, the system is equivalent to a set of normal but {\it not} independent emitters, i.e., $[\hat {\cal A}_\nu,\hat{\cal A}_{\nu'}^\dag]$ for $\nu\neq \nu'$ is generally non-zero (e.g., $[\hat {\cal A}_1,\hat{\cal A}_{2}^\dag]=e^{i k_0 (x_{21}-x_{11})}[\hat {A}_1,\hat{A}_{2}^\dag]=0$ but $[\hat {\cal A}_2,\hat{\cal A}_{4}^\dag]=e^{i k_0 (x_{22}-x_{21})}[\hat {A}_2,\hat{A}_{2}^\dag]\neq 0$). (c): Transformation from indexing (b) to (a) is described by the pair of index functions $j={\cal J}_\nu$ and $\ell={\cal L}_\nu$. These and the inverse function can be represented through the plotted diagram, where values of $\nu$ (in red) label the black dots. The Cartesian coordinates of each dot indicate the corresponding pair $(j,\ell)$. The diagram thus encodes the coupling points topology. (d): Implementation of the setup in (a) via a looped unidirectional waveguide.}\label{system}
\end{figure}

Let $S$ be a system made out of $N_e$ quantum ``emitters'' of frequency $\omega_0$ and associated ladder operators $\hat A_{j}$, $\hat A_{j}^\dag$ for $j=1,...,N_e$. The statistical nature of these operators is left unspecified, hence in particular each emitter could be a harmonic oscillator or a pseudo-spin (linear and non-linear, respectively).  The emitters are weakly coupled to a unidirectional bosonic field with normal-mode ladder operators $\hat b_\omega$, $\hat b_\omega^\dag$ such that $[\hat b_\omega,\hat b_{\omega'}]=[\hat b_\omega^\dag,\hat b_{\omega'}^\dag]=0$ and $[\hat b_\omega,\hat b_{\omega'}^\dag]=\delta(\omega-\omega')$. The $j$th emitter interacts with the field at ${\cal N}_j$ distinct coupling points. For ${\cal N}_j=1$ we retrieve the standard local coupling and the emitter is called ``normal'' [see \fig\ref{system}(a)]. Instead, if ${\cal N}_j\ge 2$, the coupling is multi-local and the emitter is dubbed ``giant'' [see \figs\ref{system}(a) and (d)]. The spatial coordinate of the $\ell$th coupling point of the $j$th emitter is $x_{j\ell}$ (the field is along the $x$-axis). Under the usual rotating-wave approximation (RWA) and assuming white coupling, the total Hamiltonian reads (we set $\hbar=1$)
\begin{eqnarray}
\hat H=\hat H_{S}+\hat H_{f}+\hat V\,\label{Htot}
\end{eqnarray}
\begin{align}
\hat H_{S}&=\sum_{j=1}^{N_e} \omega_0 \,\hat A_{j}^\dag\hat A_{j}\,,\,\,\,\,\hat{H}_{f}= \int \!d\omega\, (\omega_0+\omega)\, {\hat b}^\dagger_\omega {\hat b}_\omega \,, \label{Hf}\\
\hat V&=\sum_{j=1}^{N_e}\sum_{\ell=1}^{{\cal N}_j}\,\sqrt{\tfrac{\gamma}{2\pi}}\,e^{i \omega_0 \tau_{j\ell}}\!\int d\omega\,e^{i \omega \tau_{j\ell}}\hat A_{j}^\dag {\hat b}_\omega+ {\rm H.c.}\,,\label{V}
\end{align}
where all integrals run over the entire real axis compatibly with the RWA. 
Here, $\tau_{j\ell}=x_{j\ell}/v$ is the coordinate in the time domain of each coupling point (the field dispersion law is $\omega=v k$). Note that here $\omega$ are frequencies measured from the emitters' energy $\omega_{0}$ (i.e., detunings in fact).
We also point out that each coupling point has an associated position-dependent phase factor $e^{i \omega_0\tau_{j\ell} }$, which can be equally written in the space domain as $e^{i k_0x _{j\ell} }$ with $k_0=\omega_0/v$. 

Instead of $\omega$-dependent normal modes, the field can be equivalently represented in terms of time modes with ladder operators
\begin{align}
\hat b _t  = \tfrac{1}{\sqrt{2 \pi}} \!\int \!d\omega  \,{\hat b}_\omega e^{-i \omega t}\,,\label{tm}
\end{align}
fulfilling bosonic commutation rules
\begin{equation}
[\hat b_t, \hat b_{t'}^\dag]=\delta(t-t'),\,\, [\hat b_t,\hat b_{t'}]=[\hat b_t^\dag,\hat b_{t'}^\dag]=0\,\,.\label{rt-comm}
\end{equation}

\subsection{Interaction picture and relabeling}\label{sub-ip}

Passing to the interaction picture with respect to $\hat H_0=\hat H_{S}+\hat H_f$ transforms ladder operators as $\hat A_{j}{\rightarrow} \hat A_{j} e^{-i\omega_0 t}$ and $\hat b_{\omega}\rightarrow \hat b_{\omega} e^{-i(\omega_{0}+\omega) t}$ so that the joint emitter-field state $\sigma$ now evolves as $\dot \sigma=-i \,[\hat V_t ,\sigma]$ with
\begin{equation}
\hat V_t=\sqrt{\gamma}\,\sum_{j,\ell}\,\hat A_{j}^\dag e^{i \omega_0 \tau_{j\ell}} \,{\hat b}_{t-\tau_{j\ell}}+ {\rm H.c.}\,\label{Vt1}
\end{equation}
Now, following \rref\cite{Hammerer2019}, it is convenient to introduce an index $\nu=1,...,{\cal N}$ labeling {all} the coupling points from left to right, i.e.,  $x_1<x_2<...<x_{\cal N}$ [see \fig\ref{system}(b)] or equivalently in the time domain  $\tau_1<\tau_2<...<\tau_{\cal N}$ (here ${\cal N}=\sum_{j=1}^{N_e} {\cal N}_j$ is the total number of coupling points).
For each coupling point $\nu$, we define a corresponding ladder operator as
\begin{equation}
\hat {\cal A}_{\nu}=\hat A_{j}\,e^{-i k_0 x_{j\ell}} \,,\label{Anu}
\end{equation} 
with $\hat A_{j}$ the ladder operator of the corresponding atom and $e^{-i k_0 x_{j\ell}}$ the corresponding phase shift. 
For instance, in the case of \fig\ref{system}(a): ${\cal A}_5=\hat A_3 e^{-i k_0 x_{32}}=\hat A_3 e^{-i \omega_0 \tau_{32}}$.  Formally, the mapping between $(j,\,\ell)$ and $\nu$ is a expressed by a pair of discrete functions $j={\cal J}_\nu$ and $\ell={\cal L}_\nu$, a diagrammatic representation of which is shown in \fig\ref{system}(c).
Note that ladder operators $\{\A_\nu\}$ with different indexes do not necessarily commute, that is $[\A_\nu,\A_{\nu'\neq \nu}^\dag]$ is {generally non-zero} [e.g., in \fig\ref{system}(b), $[{\cal A}_1, {\cal A}_3^\dag ]=0$ but $[{\cal A}_3, {\cal A}_5^\dag]\neq0$]. In this way the system could be thought as a set of ${\cal N}$ normal emitters (as many as the coupling points), which yet are {\it not} independent. Their dynamics is governed by the Hamiltonian [\cf\eq\eqref{Vt1}]
\begin{equation}
\hat V_t=\sqrt{\gamma}\,\sum_{\nu=1}^{\cal N}\,\A_{\nu}^\dag \,{\hat b}_{t-\tau_\nu}+ {\rm H.c.}\,\label{Vt}
\end{equation}

\subsection{Bidirectional field}\label{sec-bi}

For a bidirectional field, each normal frequency $\omega$ now has associated right-going and left-going modes with ladder operators $\hat b_\omega$ and ${\hat {{b}'}}_\omega$, respectively (${\hat {{b}'}}_\omega$ fulfill commutation rules analogous to $\hat b_\omega$). In the total Hamiltonian \eqref{Htot}, the field and coupling Hamitonians are replaced by
\begin{align}
\hat{H}_{f}=& \int \!d\omega\, (\omega_{0}+\omega)\, ({\hat b}^\dagger_\omega {\hat b}_\omega+{\hat {{b}'}}^\dagger_\omega {{\hat {{b}'}}}_\omega) \,, \label{Hf-rl}\\
\hat V=&\sqrt{\tfrac{\gamma}{2\pi}}\,\sum_{j,\ell}\,e^{i \omega_0 \tau_{j\ell}}\!\int d\omega\,e^{i \omega \tau_{j\ell}}\hat A_{j}^\dag\, {\hat b}_\omega\nonumber\\
&+\sqrt{\tfrac{\gamma'}{2\pi}}\,\sum_{j,\ell}\,e^{-i \omega_0 \tau_{j\ell}}\!\int d\omega\,e^{i \omega \tau_{j\ell}}\hat A_{j}^\dag \,{{\hat {{b}'}}}_\omega+ {\rm H.c.}\,,\label{V-rl}
\end{align}
where we allowed generally different coupling strengths to right- and left-going modes so as to encompass chiral dynamics \cite{LodahlReviewNature17} (the previous unidirectional case is retrieved for $\gamma'=0$). Note the different phase factors in right-going terms compared to left-going ones. A detailed derivation of the microscopic Hamiltonian is reviewed in Appendix A.

Left-going time modes are defined analogously to \eqref{tm} as
\begin{align}
{\hat {{b}'}}\! _t  = \tfrac{1}{\sqrt{2 \pi}} \!\int \!d\omega  \,{{\hat {{b}'}}}\!_\omega \,e^{-i \omega t}\,,\label{tm2}
\end{align}
fulfilling commutation rules analogous to \eqref{rt-comm}.

Proceeding similarly to the unidirectional case leads to the interaction-picture coupling Hamiltonian [\cf\eq\eqref{Vt}]
\begin{equation}
\hat V_t=\sqrt{\gamma}\,\sum_{\nu=1}^{\cal N}\,\A_{\nu}^\dag \,{\hat b}_{t-\tau_\nu}+\sqrt{\gamma'}\,\sum_{\nu=1}^{\cal N}\,\Ap_\nu^\dag\,{{\hat {{b}'}}}\!_{t+\tau_\nu}+ {\rm H.c.}\label{Vt-rl}
\end{equation}
with $\A_{\nu}$ defined as in \eqref{Anu} and $\Ap_{\nu}$ 
\begin{equation}
\Ap_{\nu}=\hat A_{j}\,e^{i k_0 x_{j\ell}}\,,\label{Anup}
\end{equation}
where, just like in \eq\eqref{Anup}, $j={\cal J}_\nu$ and $\ell={\cal L}_\nu$ (note however the change of phase compared to $\A_{\nu}$).

\section{Summary of main results} \label{summary}

In this section, we sum up some of the main results of this work.

\subsection{Unidirectional field}

\begin{figure}
	\includegraphics[width=0.46 \textwidth]{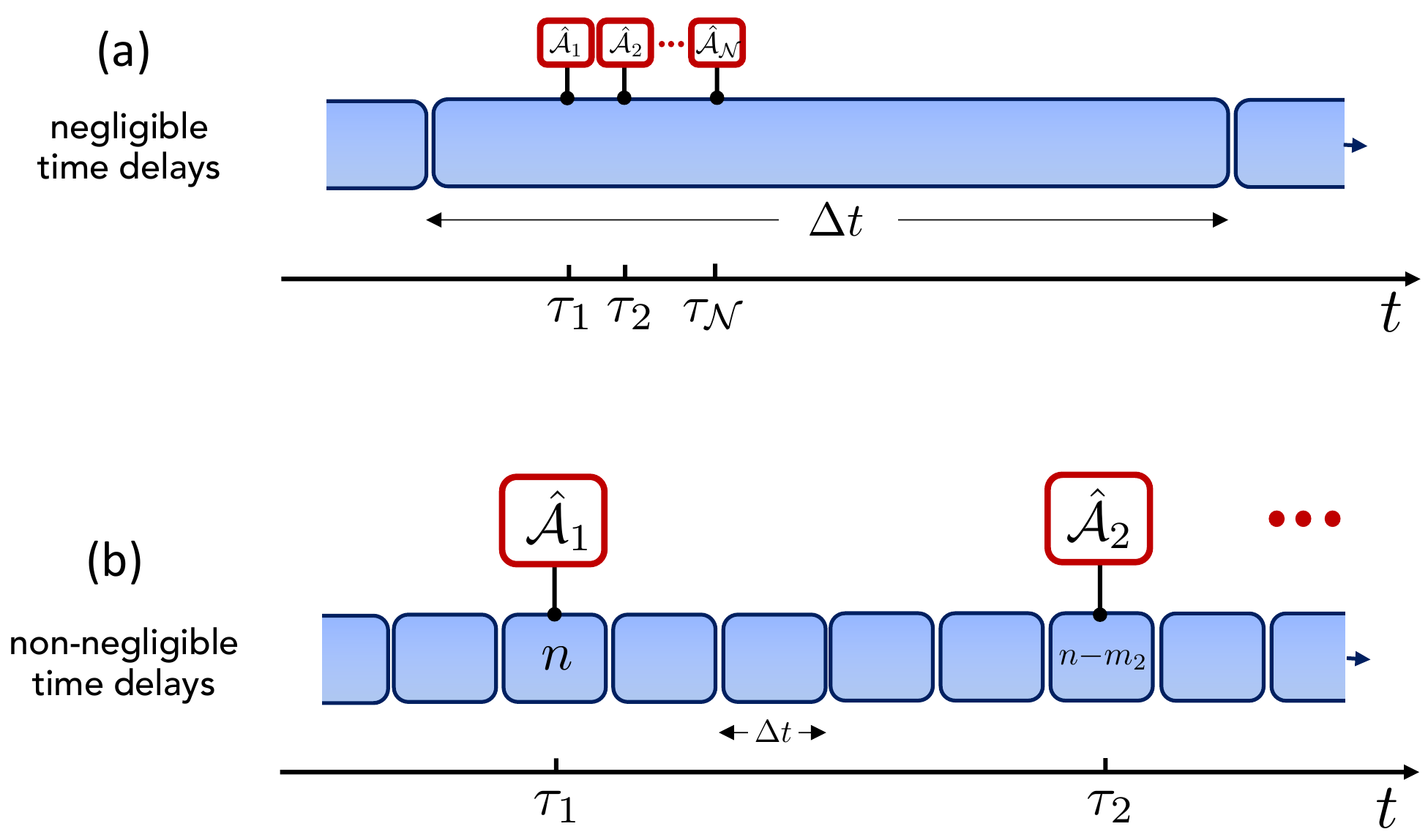}
	\caption{Effective collision model for a unidirectional field in the regimes of negligible (a) and non-negligible (b) time delays. (a): Negligible time delays, $\tau_{\nu}-\tau_{\nu-1}\ll\Delta t\ll \gamma^{-1}$ for any $\nu$. The time bin is much larger than the distance (in the time domain) between coupling points. Note, though, that so long as time delays are finite (no matter how short) the behavior is different from the ideal case of colocated coupling points: the fact that each time bin collides first with $\nu=1$, then $\nu=2$ etc.~produces the effective Hamiltonian \eqref{Hvac}. (b): Non-negligible time delays, $\Delta t\ll \tau_{\nu}-\tau_{\nu-1}\ll \gamma^{-1}$ for any $\nu$. Distinct coupling points collide with different, generally non-consecutive, time bins.}\label{delays}
\end{figure}

Let $t_n=n \Delta t$, with $n$ integer and $t_0$ the initial time, be a mesh of the time axis.
In the regime of {\it negligible time delays} defined by $t_{\cal N}-t_1\ll \Delta t\ll \gamma^{-1}$ (see \fig\ref{delays}), the propagator of the joint dynamics (in the interaction picture) is well-approximated as
\begin{equation}
\hat {\cal U}_t\simeq \prod_{n=1}^{[t/\Delta t]} \hat U_n\,\,\,\,\,{\rm with}\,\,\,\hat U_n=e^{-i \left(\hat H_{\rm vac}+\hat V_n\right)\Delta t}\,,\label{sequence} 
\end{equation}
where
\begin{eqnarray}
\hat H_{\rm vac}&=&i\tfrac{\gamma}{2}\sum_{\nu>\nu'}\left(\A_{\nu'}^\dag \A_{\nu}-\A_{\nu}^\dag \A_{\nu'}\right)\,,\label{Hvac}\\
\hat V_{n}&=&\sqrt{\tfrac{\gamma}{\Delta t}}\,\left(\A^\dag  \hat b_n+{\rm H.c.}\right)\,\label{Vn}
\end{eqnarray} 
(note the characteristic $1/\sqrt{\Delta t}$ dependence of the coupling strength).
Here, $\A$ is the collective emitters' operator $\A=\sum_\nu \A_\nu$, while
\begin{equation}
\hat b_n=\tfrac{1}{\sqrt{\Delta t}}\int_{t_{n-1}}^{t_n}\!dt\, \hat b_t\,\label{rn0}
\end{equation}
is the annihilation operator associated with the $n$th time bin of the field. Time-bin ladder operators fulfill standard bosonic commutation rules $[\hat b_n,\hat b_{n'}]=[\hat b^\dag_n,\hat b^\dag_{n'}]=0$ and $[\hat b_n,\hat b^\dag_{n'}]=\delta_{n,n'}$. 

Thus the dynamics effectively consists of a sequence of pairwise {\it collisions} (short interactions). During the $n$th collision, the emitters collectively couple to the $n$th field's time bin (interaction $\hat V_n$) and at the same time undergo an effective dipole-dipole interaction described by Hamiltonian $\hat H_{\rm vac}$. Note that time bins are non-interacting with each other and that the $n$th time bin interacts with the emitters only during the time interval $[t_{n-1},t_n]$ in a conveyor-belt fashion, in this respect just like the standard case of one normal emitter (see \fig\ref{sketch}). Note that the dipole-dipole Hamiltonian \eqref{Hvac} has a chiral origin: it arises because each time bin (see \fig1) collides \textit{first} with coupling point $\nu=1$, \textit{then} $\nu=2$ and so on. Indeed if all the coupling points had the same location, $\hat V_n$ would still be present but $\hat H_{\rm vac}=0$.

Let $\sigma_n=\sigma_{t=t_n}$ be the joint state of the emitters and all time bins with $\sigma_0=\rho_0\otimes\rho_f$, where $\rho_0$ ($\rho_f$) is the initial state of emitters (field).  At each collision, $\sigma_n$ evolves according to
\begin{align}
\frac{\Delta \sigma_n}{\Delta t}=&-i [\hat H_{\rm vac}+\hat V_n,\sigma_{n-1}]\nonumber\\
&+\Delta t\!\left(\hat V_n\sigma_{n-1}\hat V_n-\tfrac{1}{2}\left[\hat V_n^2,\sigma_{n-1}\right]_+\right),\label{dsigma}
\end{align}
where $\Delta \sigma_n=\sigma_{n}-\sigma_{n-1}$ and $[...,...]_+$ stands for the anti-commutator (the $\Delta t$-dependence of 2nd-order terms is only apparent since $\hat V_n\sim 1/\sqrt{\Delta t}$). 
If the initial state of the time bins corresponding to the field state $\rho_f$ is of the form $\bigotimes_n \eta_n$ (no correlations), then tracing off the field in \eq\eqref{dsigma} yields that the emitters undergo a Markovian dynamics described by
\begin{align}
\frac{\Delta \rho_n}{\Delta t}=&-i [\hat H_{\rm vac}+\langle \hat V_n\rangle,\rho_{n-1}]\nonumber\\
&+\Delta t\,{\rm Tr}_n\left\{ \hat V_n\,\rho_{n-1} \eta_n\hat V_n-\tfrac{1}{2}\left[\hat V_n^2,\rho_{n-1} \eta_n\right]_+ \right\}\label{d-ME}
\end{align}
with $\rho_n$ the state of the emitters at time $t_n$, $\Delta \rho_n=\rho_{n}-\rho_{n-1}$ and $\langle...\rangle={\rm Tr}_n\left\{...\,\eta_n\right\}$, where ${\rm Tr}_n\{\}$ is the partial trace over time bin $n$. \eq\eqref{d-ME} can always be expressed in the standard Lindblad form, ${\Delta \rho_n}/{\Delta t}=- i [\hat {\mathcal H}, \rho_{n-1}]+\sum_m {\cal D}_{\hat J_m}[\rho_{n-1}]$, with $\hat {\mathcal H}=\hat {\mathcal H}^\dag$ and
\begin{equation}
\mathcal D_{\hat J}[\rho]=\hat J \rho \hat J^\dag-\tfrac{1}{2}[\hat J^\dag\hat J, \rho]_+\label{DJ}\,\,.
\end{equation}
where $\{{\hat J}_m\}$ is a suitable collection of jump operators.
The Lindblad form is guaranteed because at each collision the emitters evolve according to a completely-positive and trace-preserving (CPT) map, $\rho_{n}=\big\langle\hat U_n\,\rho_{n-1}\hat U_n^\dag\big\rangle$.

The most general white-noise Gaussian state of the field is fully specified by the 1st and 2nd moments \cite{WisemanMilburnBook}
\begin{equation}
\langle d\hat b_t \rangle =\alpha_t \,dt\,,\,\,\,\langle d\hat b^\dag_t  \,d\hat b_t\rangle=N \,dt\,,\,\,\,\langle d\hat b_t  \,d\hat b_t\rangle=M \,dt\label{dB}\,.
\end{equation}
with $d\hat b_t=\int_t^{t+dt}\!ds\,\hat b_s$ the well-known quantum noise increment. Correspondingly, the most general Gaussian, uncorrelated state of the time bins is fully specified by the moments
\begin{equation}\label{dbn}
\langle \hat b_n \rangle =\alpha_n \,\sqrt{\Delta t},\,\langle \hat b_n^\dag   \,\hat b_{n'}\rangle=\delta_{n,n'}\,N,\,\langle \hat b_n  \,\hat b_{n'}\rangle=\delta_{n,n'}\,M \,.
\end{equation}
with $\alpha_n=\alpha_{t=t_n}$, $N\ge 0$ and $|M|^2\le N(N+1)$. 

Replacing the explicit expression of $\hat V_n$ in \eq\eqref{d-ME} using \eqref{dbn} and carrying out the continuous-time limit $\gamma\Delta t\rightarrow 0$, the discrete master equation \eqref{d-ME} is turned into the general continuous-time master equation 
\begin{align}\label{g-ME}
\frac{d \rho}{d t}=&-i \,[\hat H_{\rm vac}+\sqrt{\gamma}\, (\alpha_t^* \hat {\cal A}+{\rm H.c.}),\rho] \nonumber\\
&+\gamma (N+1){\mathcal D}_{\A}[\rho]+\gamma N{\mathcal D}_{\A^\dag}[\rho]\nonumber\\
&+\gamma \left(M(\A^\dag\,\rho\A^\dag-\tfrac{1}{2}[\A^{\dag 2},\rho]_+)+{\rm H.c.}\right)\,.
\end{align}
This can be expressed in terms of original ladder operators $\hat A_{j}$ using \eqref{Anu} and recalling $\A=\sum_\nu \A_\nu$.

\begin{figure}
	\includegraphics[width=0.45\textwidth]{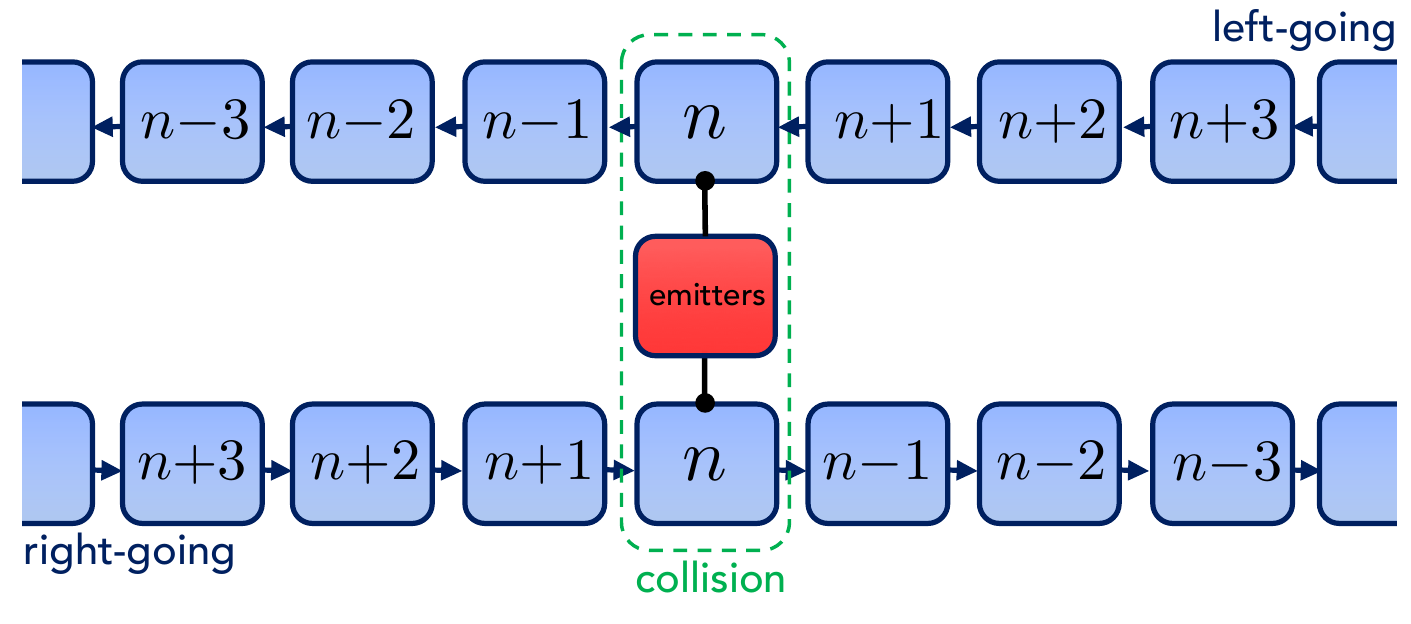}
	\caption{Collision-model description for a bidirectional field. Each time bin is now bipartite, comprising a right-going mode (bottom) and a left-going mode (top). At each collision the emitters jointly collide with the two-mode time bin according to the coupling Hamiltonian \eqref{Vn-rl} and, additionally, are subject to an internal coherent dynamics corresponding to the dipole-dipole Hamiltonian \eqref{Hvac-rl}.}\label{bipartite}
\end{figure}

\subsection{Bidirectional field}

In the case of a bidirectional field, the time bin is now bipartite (see \fig\ref{bipartite}) having associated ladder operators $\hat b_n$ [\cf\eq\eqref{rn0}] and ${\hat {{b}'}}\!_n$, the latter given by
	\begin{equation}
	{\hat {{b}'}}\!_n=\tfrac{1}{\sqrt{\Delta t}}\int_{t_{n-1}}^{t_n}\!dt\, \hat b'_t\,,\label{rtn}
	\end{equation}
while $\hat H_{\rm vac}$ and $\hat V_n$ are now generalized as
	\begin{eqnarray}
	\hat H_{\rm vac}&=&\tfrac{i}{2}\sum_{\nu>\nu'}\left( \gamma \A_{\nu'}^\dag \A_{\nu}+ \gamma' \Ap_{\nu}^\dag \Ap_{\nu'}-{\rm H.c.}\right),\label{Hvac-rl}\\
	\hat V_{n}&=&\tfrac{1}{\sqrt{\Delta t}}\,\left(\sqrt{\gamma}\,\A^\dag \hat b_n+\sqrt{\gamma'}\,\Ap^\dag {\hat {{b}'}}\!_n+{\rm H.c.}\right),\,\,\,\,\,\label{Vn-rl}
	\end{eqnarray} 
	with 
\begin{equation}
	\A=\sum_\nu \A_\nu\,,\,\,\,\Ap=\sum_\nu \Ap_\nu\,.\label{coll}
\end{equation}
Inside brackets of \eqref{Hvac-rl}, note that the second term has swapped subscripts compared to the first. This is due to the opposite interaction time ordering for left- and right-going modes.

Accordingly, the dissipator in the Lindblad master equation \eqref{g-ME} now naturally splits into a pair of analogous contributions: one featuring operators $\A_\nu$'s and moments of right-going field modes ($\alpha_{t}$, $N$, $M$) and another one involving $\Ap_\nu$'s and left-going-mode moments ($\alpha'_{t}$, $N'$, $M'$). The latter moments are defined analogously to \eqref{dB} with $\hat b_t\rightarrow {\hat {{b}'}}_t$. This leads to the master equation
\begin{align}\label{g-ME-rl}
\frac{d \rho}{d t}=&-i \,[\hat H_{\rm vac}+\sqrt{\gamma}\, (\alpha_{t}^* \hat {\cal A}+{\alpha'}_{t}^* \Ap+{\rm H.c.}),\rho] \nonumber\\
&+\gamma (N+1){\mathcal D}_{\A}[\rho]+\gamma N{\mathcal D}_{\A^\dag}[\rho]\nonumber\\
&+\gamma' (N'+1){\mathcal D}_{\Ap}[\rho]+\gamma' N'{\mathcal D}_{\Ap^\dag}[\rho]\nonumber\\
&+\gamma \left(M(\A^\dag\rho\,\A^\dag-\tfrac{1}{2}[\A^{\dag 2},\rho]_+)+{\rm H.c.}\right)\nonumber\\
&+\gamma' \left(M'(\Ap^\dag\rho\,\Ap^\dag-\tfrac{1}{2}[\Ap^{\dag 2},\rho]_+)+{\rm H.c.}\right)\,.
\end{align}
where we recall \eqs\eqref{DJ} and \eqref{Hvac-rl}.
This can be expressed in terms of original ladder operators $\hat A_{j}$ through \eqref{Anu}, \eqref{Anup} and 	\eqref{coll}. Master equation \eqref{g-ME} for a unidirectional field is retrieved for $\gamma'=0$.

\subsection{Photodetection and quantum trajectories} \label{sec-jump-summ}

For a {\it unidirectional field}, photodetection translates into measuring each time bin right after its collision with $S$ in a selected basis $\{\ket{k}_n\}$ (defining the photodetection scheme). 
For time bins initially in state $\bigotimes_n \!\ket{\chi_n}$ (thus $\eta_n{=}\ket{\chi_n}\!\bra{\chi_n}$) and negligible time delays, the (unnormalized) state of $S$ after a specific sequence of measurement outcomes $\{k_1,...,k_n\}$ is 
\begin{equation}
\tilde \rho_n=\hat {K}_{k_n}\cdots\hat {K}_{k_1}\,\rho_0\, \hat {K}_{k_1}^\dag\cdots \hat {K}_{k_n}^\dag\label{cond-sum}\,,
\end{equation}
the associated probability being $p_{k_1\cdots \,k_n}={\rm Tr}_S\{{\tilde\rho_n}\}$ and with each Kraus operator given by
\begin{equation}
\hat {K}_{k_m}=\bra{k_m}\!\hat{ U}_m\ket{\chi_m}\,. \label{kraus-sum}
\end{equation}
A measurement on time bin $n$ with outcome $k$ (at the end of the $n$th collision) thus projects $S$ into the (unnormalized) state $\tilde\rho_n=\hat K_k \,\rho_{n-1} \,\hat K_k^\dag$ with probability $p_k={\rm Tr}_S\{\tilde\rho_n\}$, defining the conditional dynamics. Summing over all possible outcomes yields the CPT map $\rho_n={\cal E}[\rho_{n-1}]=\sum_k \hat K_k \,\rho_{n-1} \,\hat K_k^\dag $, defining the unconditional dynamics.

A (pure) time-bin state generally depends itself on $\sqrt{\Delta t}$. We consider those states such that to the lowest order in $\sqrt{\Delta t}$ read
\begin{align}
|\chi_n\rangle\simeq |0_n\rangle+|\chi_n^{(1)}\rangle\sqrt{\Delta t}+|\chi_n^{(2)}\rangle\,\Delta t\,\label{chi-exp-sum}
\end{align}
with $|\chi_n^{(j)}\rangle$ such that $\langle \chi_n|\chi_n\rangle=1+O(\Delta t)$ ($|\kappa_n\rangle$ with $\kappa_n=0,1,...$ denote the time-bin Fock states). Plugging this and $\hat U_n=e^{-i \left(\hat H_{\rm vac}+\hat V_n\right)}$ into $\rho_n={\cal E}[\rho_{n-1}]$ and dropping high-order terms eventually leads to the master equation
\begin{align}
\frac{\Delta \rho_n}{\Delta t}=- i\, [\hat H_{\rm eff},\rho_{n-1}]+\sum_{k}  {\cal D}_{\hat J_k}[\rho_{n-1}]\,\label{me-jumps-sum}
\end{align}
with the effective Hamiltonian $\hat H_{\rm eff}$ and jump operators ${\hat J_k}$ given by
\begin{align}
\hat H_{\rm eff}&=\hat H_{\rm vac}{+}\left(\tfrac{1}{2}\sqrt{\gamma}\,\langle 0_n|{\hat b}_n|\chi_n^{(1)}\rangle \,\hat {\cal A}^\dag+{\rm H.c.}\right)\,,\label{Heff-sum}\\
{\hat J_k}&=\langle k_n|\chi_n^{(1)}\rangle-i  \sqrt{\gamma}\, \langle k_n|1_n\rangle\,\A \,\,.\label{Jk-sum}
\end{align}
This in fact defines an unraveling of master equation \eqref{g-ME} [which is indeed equivalent to \eqref{me-jumps-sum}] corresponding to the photodetection scheme $\{\ket{k_n}\}$ in the case of a unidirectional field. 
\\
\\
For a {\it coherent-state wavepacket} of amplitude $\xi_t$ (in the time domain) \cite{note-xi}, $|\xi\rangle=e\,^{\int \!{d}t\,\left(\xi_t \hat b^\dag_t-\xi_t^*  \hat b_t\right)}\,|0\rangle$ (with $|0\rangle$ the field vacuum), the corresponding time-bin state is
\begin{equation}
\ket{\chi_n}= e^{\xi_n\sqrt{\Delta t}\,\hat b_n^\dag-\xi_n^*\sqrt{\Delta t}\,  \hat b_n}\ket{0_n}\,\label{chin-coh-sum}
\end{equation}
with $\xi_n=\xi_{t=t_n}$. Hence, $|\chi_n^{(1)}\rangle=\xi_n\ket{1_n}$. In the case of photon counting, $\{\ket{k_n}\}$ are the Fock states. The effective Hamiltonian and the only surviving jump operator are thus immediately calculated as
\begin{align}
\hat H_{\rm eff}=\!\hat H_{\rm vac}{+}\tfrac{1}{2}\!\sqrt{\gamma\,}\,(\xi_n \, \A^\dag{+}{\rm H.c.})\,,\hat J_1=\xi_n {-}i\sqrt{\gamma}\,\A\,.\label{HJ-coh}
\end{align}
The continuous-time limit expressions are simply obtained by replacing $\xi_n \rightarrow \xi_t$.
\\
\\
For a {\it bidirectional field}, photodetection consists in measuring both the right- and left-going time bins (see \fig\ref{bipartite}) in a basis $\ket{k,k'}$. A measurement outcome $k,k'$ is now described by the Kraus operator [\cf\eq\eqref{kraus-sum}] 
\begin{equation}
\hat {K}_{k,k'}=\bra{k,k'}\!\hat{ U}_m\ket{\chi_m,\chi'_m}
\end{equation} 
with $\ket{\chi_m}$ ($\ket{\chi'_m}$) the initial state of the right-going (left-going) time bin and $\hat U_n=e^{-i \left(\hat H_{\rm vac}+\hat V_n\right)}$ with $\hat H_{\rm vac}$ and $\hat V_n$ now given by \eqref{Hvac-rl} and \eqref{Vn-rl}.

The effective Hamiltonian and jump operators are given by [\cf\eqs\eqref{Heff-sum} and \eqref{Jk-sum}]
\begin{align}
\hat{H}_{\rm eff} =\,& 
\hat{H}_{\rm vac} \nonumber 
+ \tfrac{1}{2}\left(\sqrt{\gamma}   \bra{0} \hat{b}_n \ket{\chi^{(1)} }  \A^\dag
\right.\\&
\left.\,\,\,\,\,\,\,\,\,\,\,\,\,\,\,\,\,\,\,+\sqrt{\gamma'}  \bra{0} \hat{b'}_n \ket{ \chi'^{(1)}} \Ap^\dag + \rm H.c. 
\right)\,, \label{H_2w}\\
\hat{J}_{k,k'}=\,&\la k | 0\ra \la k' | \chi'^{(1)} \ra+  \la k' | 0 \ra \la k| \chi^{(1)} \ra\nonumber \\&\,\,\,\,\,-i  \left(\sqrt{\gamma} \,\bra{k'} 0\rangle \bra{k}1\rangle \A  \nonumber\right. \\&\left.\,\,\,\,\,\,\,\,\,\,\,\,\,\,\,\,\,+ \sqrt{\gamma'} \bra{k} 0\rangle\bra{k'} 1\ra\Ap\right)\,. \label{J_2w}
\end{align}
Note that the unraveling defined by $\hat{H}_{\rm eff}$ and $\hat{J}_{k,k'}$  can also be exploited as an effective recipe to numerically solve master equation \eqref{g-ME-rl} especially when $N_e$ is large, which generalizes to giant emitters quantum-jump methods employed for normal emitters (see, e.g., \rrefs\cite{mahmoodian2019dynamics,ManzoniNatComm17}).

\section{Examples of master equations and decoherence-free Hamiltonians} \label{examples}

The aim of this section is to illustrate how \eqref{g-ME-rl} encompasses and generalizes various quantum optics and waveguide QED master equations with a special focus on giant atoms and decoherence-free Hamiltonians. As such, it could be skipped by a reader solely interested in the collision-model derivation.

For a single normal emitter, $N_e={\mathcal N}=1$, $\A_1\equiv \hat A$ (setting $x_1=\tau_1=0$) and $\hat H_{\rm vac}=0$. Thus ME \eqref{g-ME} [or \eqref{g-ME-rl} for $\gamma'=0$] reduces to the well-known general ME of quantum optics 
for a point-like atom or harmonic oscillator \cite{WisemanMilburnBook}.

For a pair of normal emitters coupled to a unidirectional field, we have: $N_e={\cal N}=2$ and $\A_\nu\equiv e^{-i\omega_0 \tau_\nu}\hat A_\nu=e^{-ik_0 x_\nu}\hat A_\nu$ with $\nu=1,2$ (operators with different $\nu$'s in this case commute). Hence, $\hat {H}_{\rm vac}=i\tfrac{\gamma}{2}(\A_{1}^\dag \A_{2}-\A_{2}^\dag \A_{1})$ and $\A=\A_1 + \A_2$ so that for $\alpha_t=N=M=0$ (vacuum) \eqref{g-ME} [or \eqref{g-ME-rl} for $\gamma'=0$] reduces to the well-known ME of a pair of cascaded emitters in vacuum \cite{GardinerPRLcascaded,CarmichaelPRLcascaded}.

\begin{figure}
	\includegraphics[width=0.47\textwidth]{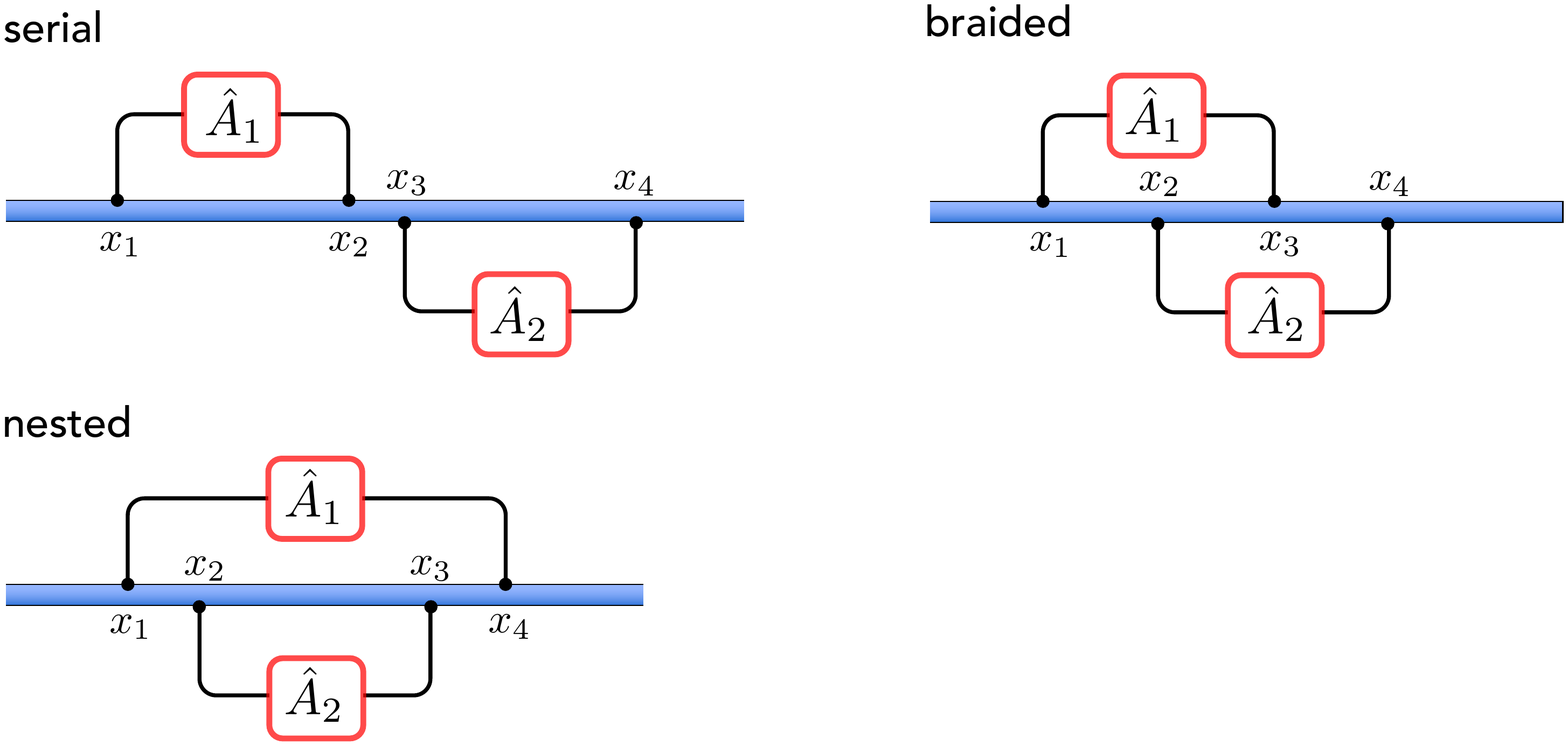}
	\caption{Possible topologies for the pattern of coupling points of two giant emitters: serial, nested and braided.}\label{topo}
\end{figure}
For $N_e={\cal N}$ normal emitters coupled to a bidirectional field (such that $\gamma'=\gamma=\Gamma/2$) \eq\eqref{g-ME-rl} reduces to 
\begin{align}
\dot \rho=&\! -i\tfrac{\Gamma}{2} \!\sum_{i \neq j}   \sin(k_0 x_{ij}^{-})[\hat A_{i}^\dag \hat A_{j},\rho]\nonumber\\
&+\! \Gamma (N{+}1)\!\sum_{ij} \cos(k_0 x_{ij}^{-}) 
\!\left(\hat{A}_i \rho \hat{A}_j^\dag {-} \tfrac{1}{2}[\hat{A}_j^\dag \hat{A}_i, \rho]_{+}\right) \nonumber
\\
&+\!\Gamma N \!\sum_{ij}  \cos(k_0 x_{ij}^{-})\!
\left(\hat{A}_i^\dag \rho \hat{A}_j {-} \tfrac{1}{2}[\hat{A}_j \hat{A}^\dag_i, \rho]_{+}\right)\nonumber
\\&
+\!\Gamma\!\sum_{ij}  \cos(k_0 x_{ij}^{+})
\!\left(M (\hat{A}_i^\dag \rho \hat{A}^\dag_j {-} \tfrac{1}{2}[\hat{A}^\dag_j \hat{A}^\dag_i, \rho]_{+}) {+} \rm H.c.\right)\label{zub}
\end{align}
with $x_{ij}^{\pm} = x_j \pm x_i$ and where we used that $\hat {\cal A}_j=\hat A_j e^{-i k_0 x_j}$, $\Ap_j=\hat A_j e^{i k_0 x_j}$ (atom and coupling-point indexes coincide).  For $N=N'=\sinh^2(|\xi|)$ and $M=M'=e^{-i\theta}\sinh(|\xi|) \cosh(|\xi|)$ \eq\eqref{zub} reduces to the master equation found through standard methods in \rref\cite{ZubairyPRA2018} ($\xi = |\xi| e^{-i \theta}$ is the squeezing parameter, where $\theta$ can include contributions depending on the distance from the source).

For zero squeezing, $\xi=N=M=0$, \eq\eqref{zub} reduces to the standard waveguide-QED master equation of a set of atoms \cite{TudelaPRL11,ChangNJP12}. 

Thus \eq\eqref{g-ME-rl} generalizes the squeezed-bath master equation to giant emitters.

For a single giant emitter with two coupling points in a bidirectional waveguide: $\gamma'=\gamma=\Gamma/2$ (with $\Gamma$ the total decay rate), $N_e=1$, ${\cal N}=2$, $\A_1=\Ap_1=\hat A_1=\hat A$, $\A_2= e^{-i\varphi}\hat A$, $\Ap_2= e^{i \varphi} \hat A$, where we set $x_1=\tau_1=0$ and $\varphi=k_0 x_2=\omega_0 \tau_2$ ($\hat A$ could be a spin-1/2 or bosonic ladder operator). The collective operators \eqref{coll} thus read 
\begin{equation}
\A=(1+e^{-i\varphi})\,\hat A\,,\,\,\,\Ap=(1+e^{+i\varphi})\,\hat A\,. \label{coll-giant}
\end{equation}

Plugging these into \eq\eqref{g-ME-rl}, for $\alpha_t=\alpha'_t=N=N'=M=M'=0$ we retrieve the vacuum master equation \cite{TufarelliPRA13,KockumPRL2018}
\begin{equation}
\dot \rho = - i \,\tfrac{\Gamma}{2} \sin \varphi\, [\hat{A}^\dag \hat{A} , \rho] + \Gamma (1 + \cos\varphi) {\cal D}_{\hat A}[\rho]\,.
\end{equation}

For a pair of giant emitters with two coupling points each and a bidirectional waveguide: $\gamma'=\gamma=\Gamma/2$, $N_e=2$, ${\cal N}=4$. The $\A_\nu$'s and $\Ap_{\nu}$'s depend on the pattern of coupling points, for which three different  topologies are possible: serial, nested and braided (see \fig\ref{topo}). Setting $\varphi_\nu=k_0 x_{\nu}=\omega_{0}\tau_\nu$ and as usual $x_1=\tau_1=0$, in the braided configuration in particular one gets $\A_1=\hat A_1$, $\A_2=\hat A_2 e^{-i \varphi_2}$, $\A_3=\hat A_1 e^{-i \varphi_3}$ and $\A_4=\hat A_2 e^{-i \varphi_4}$. Hence [\cf\eq\eqref{coll}], 
\begin{equation}
\A=(1+e^{-i \varphi_3})\hat A_1+(e^{-i \varphi_2}+e^{-i \varphi_4})\hat A_2 \label{coll2}\,,
\end{equation}
while $\Ap$ has an analogous expression with $\varphi_\nu\rightarrow -\varphi_\nu$. Plugging these into \eqref{g-ME-rl}, for $\varphi_\nu=\nu \varphi$ (uniform spacings) and the field {vacuum} state, one gets
\begin{align}
\dot \rho = &- i \tfrac{\Gamma}{2}\, (3 \sin\varphi {+} \sin 3\varphi) \left[\hat{A}_2^\dag \hat{A}_1{+}\hat{A}_1^\dag \hat{A}_2 , \rho \right] \nonumber
\\&+ 2 \Gamma\, (1 {+} \cos 2\varphi) \left(\mathcal{D}_{\hat{A}_1} [\rho ]{+} \mathcal{D}_{\hat{A}_2} [\rho] \right) \nonumber
\\&+ \Gamma\,(3 \cos\varphi{+} \cos3\varphi) \sum_{i \neq j} \left(\hat{A}_i \rho \hat{A}_j^\dag - \tfrac{1}{2}[\hat{A}_j^\dag \hat{A}_i,\rho]_{+}\right)\,,
\label{braided}
\end{align}
which was derived through the SLH formalism in \rref\cite{KockumPRL2018} alongside other master equations for different configurations and number of atoms [these can all be retrieved from \eqref{g-ME-rl} likewise].

\subsection{Decoherence-free Hamiltonians with giant atoms}\label{section-DF}

A major appeal of giant emitters is that they allow to implement decoherence-free many-body Hamiltonians. A paradigmatic instance is the braided configuration in \fig\ref{topo}. By adjusting a $\pi$-phase shift between the coupling points of the same emitter, e.g., setting $\varphi=\pi/2$, all the dissipative terms in \eq\eqref{braided} vanish but the Hamiltonian $\hat H_{\rm vac}$, which effectively seeds a  dissipationless coherent interaction \cite{KockumPRL2018}. 

In the collisional picture this phenomenon can be predicted without working out the master equation, making clear at once that it occurs regardless of the field state [thus being not limited to the vacuum state assumed in the derivation of \eq\eqref{braided}]. Indeed, the condition that collective operators \eqref{coll} vanish,
\begin{equation}
\A=\Ap=0\label{DF}
\end{equation}
(or just $\A=0$ with a unidirectional field), guarantees that the joint emitters-field propagator reduces to $\hat {\cal U}_t=\exp({-i {\hat H}_{\rm vac}t})$. This is because \eqref{DF} effectively decouples the emitters from the field time bins in light of \eqs\eqref{sequence}, \eqref{Hvac-rl} and \eqref{Vn-rl}, thus inhibiting dissipation. Having giant emitters is clearly indispensable since for normal emitters there is no way for $\A$ and $\Ap$ to identically vanish in the entire Hilbert space.
The question is now whether or not \eqref{DF} yields in addition a null $\hat H_{\rm vac}$ (if so no evolution takes place). For a giant atom [\cf\eq\eqref{coll-giant}], the condition $\A=\Ap=0$ holds for $\varphi=(2n{+}1)\pi$ which will also entail $\hat H_{\rm vac}=0$. For two giant atoms, the collective operators vanish for any $\pi$-phase shift between the coupling points of the same emitter [\cf\eq\eqref{coll2}]. Using \eqref{Hvac-rl}, one can check that this always yields $\hat H_{\rm vac}=0$ in the serial and nested topologies (see \fig\ref{topo}) whereas in the braided one $\hat H_{\rm vac}$ can be non-zero (for a comprehensive analysis we point the reader to \rref\cite{CCC}).

In the collisional picture, occurrence of $\hat H_{\rm vac}\neq 0$ with zero decoherence means that each time bin ends up uncorrelated with the emitters as the collision is complete. Notwithstanding,  during the collision, it mediates a crosstalk between the emitters which thus get correlated with one another.

\section{Collision model derivation}\label{derivation}

In this section, we address the derivation of the collision model for a unidirectional field (the generalization to the bidirectional case is presented in Section \ref{CM-bi}). 

Two regimes stand out:
\begin{enumerate}
	\item Negligible time delays: $\tau_{\cal N}-\tau_{1} \ll \gamma^{-1}$ (hence $\tau_{\cal N}-\tau_{1} $ can be replaced with $\tau_\nu-\tau_{\nu-1}$ for all $\nu$'s);
	\item Non-negligible time delays: significant value of $\gamma(\tau_\nu-\tau_{\nu-1})$ for any $\nu$ (say of the order of $\sim 0.1$ or larger).
\end{enumerate}
Note that regime (1) is often dubbed ``Markovian''. Strictly speaking, this is an abuse of language relying on the fact that for many typical field states (such as vacuum, thermal, coherent or broadband squeezed states) dynamics in regime (1) are Markovian and described by a Lindblad master equation. This is not necessarily the case, though, with more general field states such as single-photon wavepackets, even for a single coupling point \cite{GheriChapter05}. 
Intermediate regimes between (1) and (2) are of course possible, but these can be described as a combination of (1) and (2). 

Most of the present section concerns the regime of {\it negligible time delays} (1) (our main focus in this work), which still occurs in the vast majority of experimental setups (see \eg \rref\cite{LalumierePRA13} for a discussion on circuit-QED systems). Nevertheless, we begin with some general considerations and properties common to both regimes. 

Consider a time mesh defined by $t_n=n \Delta t$ with $n=0,1,...$ integer and $\Delta t$ the time step (later on this will be interpreted as the collision time).
In the interaction picture (see Section \ref{sub-ip}), the propagator $\U_t$ can be decomposed as \cite{note1}
\begin{equation}
\U_t=\hat {\mathcal{T}} \,e^{-i\int_{t_{0}}^{t} ds\, \V(s)}=\prod_{n=1}^{[t/\Delta t]} \hat U_n\,\,,\label{Ut2}
\end{equation}
with $\V(s)$ given in \eq\eqref{Vt1} and $\hat {\mathcal{T}}$ the usual time-ordering operator, and where each unitary $\hat U_n$ describes the evolution in the time interval $t\in [t_{n-1},t_n]$
\begin{align}\label{Magnus}
\hat U_n=\hat {\mathcal{T}}\,e^{-i\int_{t_{n-1}}^{t_n} ds\, \V_s}\,.
\end{align}
This discretization of the joint dynamics underpins the collision-model description (in any regime). Throughout, we will consider a time step much shorter than the characteristic interaction time, i.e., $\Delta t\ll \gamma^{-1}$. Accordingly, we apply Magnus expansion \cite{Magnus1954} and approximate \eqref{Magnus} up to second order in $\Delta t$ as
\begin{align}
\hat U_n\simeq \openone -i \,(\HH_n^{(0)}  + \HH_n^{(1)} ) \Delta t -\tfrac{1}{2} (\HH_n^{(0)})^2 \Delta t^2\label{Un-app}
\end{align}
with $\openone$ the identity operator and 
\begin{eqnarray}
\HH_n^{(0)}&=&\tfrac{1}{\Delta t}\!\int_{t_{n-1}}^{t_n} \!ds \,\V_s\,,\label{Mag1}\\
\HH_n^{(1)}&=&\tfrac{i}{2\Delta t}\!\int_{t_{n-1}}^{t_n} \!ds \int_{t_{n-1}}^{s} \!ds' \,[\V_{s'},\V_{s}]\label{Mag2}
\end{eqnarray}
(note that $\HH_n^{(1)}$ is Hermitian).

Using \eqref{Vt}, $\HH_n^{(0)}$ more explicitly reads 
\begin{align}
\HH_n^{(0)}
&=
\frac{1}{\Delta t}\int_{t_{n-1}}^{t_n}ds\sqrt{\gamma}\,\sum_\nu\,\A_{\nu} \,{\hat b}_{s-\tau_\nu}^\dag+ {\rm H.c.}\nonumber
\\
&=
\sqrt{\tfrac{\gamma}{\Delta t}}\,\sum_\nu\A_{\nu}\left(\tfrac{1}{\sqrt \Delta t}\!\int_{t_{n-1}-\tau_\nu}^{t_n-\tau_\nu} \!ds\,\,  \hat b^\dag_{s}\right)\!+{\rm H.c.}\label{Mag1-1}\,,
\end{align}
while $\HH_n^{(1)}$ is the sum of three terms
\begin{equation}
\HH_n^{(1)}=\HH_{\rm vac}^{(1)}+\HH_{\rm th}^{(1)}+\HH_{\rm sq}^{(1)} \label{HH1}
\end{equation}
with \cite{note-sub}
\begin{eqnarray} 
\HH_{\rm vac}^{(1)}=\,&&i\tfrac{\gamma}{2\Delta t}\!\sum_{\nu\nu'}\!\A_{\nu'}^\dag \A_{\nu}\!\int_{t_{n-1}}^{t_n} \!\!ds\! \int_{t_{n-1}}^{s}\!\! \!ds'\,[ {\hat b}_{s'-\tau_{\nu'}}{,}{\hat b}^\dag_{s-\tau_{\nu}}]\nonumber\\
&&+{\rm H.c.}\,,\label{H1vac}\\
\HH_{\rm th}^{(1)}=\,&&i\tfrac{\gamma}{2\Delta t}\sum_{\nu\nu'}\,[\A_{\nu'}^\dag,\A_{\nu}]\int_{t_{n-1}}^{t_n} ds \int_{t_{n-1}}^{s} ds'\,\,  {\hat b}^\dag_{s-\tau_{\nu}}{\hat b}_{s'-\tau_{\nu'}}\nonumber\\
&&+{\rm H.c.}\label{H1th}\,,\\
\HH_{\rm sq}^{(1)}=\,&&i\tfrac{\gamma}{2\Delta t}\sum_{\nu\nu'}\,[\A_{\nu'},\A_{\nu}]\int_{t_{n-1}}^{t_n} ds \int_{t_{n-1}}^{s} ds'\,\,  {\hat b}^\dag_{s-\tau_\nu}{\hat b}^\dag_{s'-\tau_{\nu'}}\nonumber\\
&&+{\rm H.c.}\label{H1sq}\,
\end{eqnarray}

\subsection{Negligible time delays}\label{negli}

When time delays are negligible we can coarse grain the dynamics over a time scale defined by $\Delta t$ such that
\begin{equation}
\tau_{\cal N}-\tau_1\ll \Delta t\ll\gamma^{-1}\label{hie}\,,
\end{equation}
meaning that the overall length of the coupling points array (hence the distance between any pair $\tau_\nu-\tau_{\nu'}$) is negligible compared to the time step defining the time scale [see \fig\ref{delays}(a)].

We can take advantage of \eqref{hie} and obtain approximated expressions of $\HH_n^{(0)}$ and $\HH_n^{(1)}$. As for $\HH_n^{(0)}$, the lower and upper limits of integration of each integral appearing in \eqref{Mag1-1} can be approximated as $t_{n-1}-\tau_\nu\simeq t_{n-1}$ and $t_{n}-\tau_\nu\simeq t_{n}$ so that (we set $\tau_1=0$ throughout)
\begin{align}\label{neg-delay-approx}
  \int_{t_{n-1}-\tau_\nu}^{t_n-\tau_\nu} \!ds\,\,  \hat b_{s}
  &\simeq
  \int_{t_{n-1}}^{t_n} \!ds\,\,  \hat b_{s}=\sqrt{\Delta t}\,\hat b_n\,,
\end{align}
where we defined the $\hat b_n$'s as \eqref{rn0}. It is easily checked that the commutation rules for the $\hat b_t$'s [\cf\eq\eqref{rt-comm}] entail $[\hat b_n,\hat b_m^\dag]=\delta_{nm}$ and $[\hat b_n,\hat b_m]=[\hat b^\dag_n,\hat b^\dag_m]=0$. Thus the $\hat b_n$'s define a discrete collection of bosonic modes, which we will usually refer to in the remainder as ``time-bin modes`` or at times simply as ``time bins''. Thus \eqref{Mag1-1} in the present regime reduces to
\begin{equation}
\HH^{(0)}_n\simeq \hat V_n=\sqrt{\tfrac{\gamma}{\Delta t}}\,(\A \,\hat b_n^\dag+{\rm H.c.})\,,\label{HH0-Vn}
\end{equation}
where $\A=\sum_\nu \hat A_\nu$ is a collective operator of the emitters. Note the characteristic scaling $\sim \Delta t^{-1/2}$ of the emitter-(time bin) coupling strength, which is a hallmark of CMs \cite{ciccarelloCollision2017}.

In line with approximation \eqref{neg-delay-approx}, in \eqs\eqref{H1th} and \eqref{H1sq} all time delays can be neglected replacing $s-\tau_\nu$ ($s'-\tau_{\nu'}$) with $s$ ($s'$). Based on this, In Appendix \ref{app1} we show that both $\HH^{(1)}_{\rm th}$ and $\HH^{(1)}_{\rm sq}$ can be {\it neglected} (note that Appendix \ref{app1} refers to Section \ref{sec-bnk} to be discussed shortly).

\begin{figure}[t!]
	\begin{center}
		\includegraphics[width=0.22\textwidth]{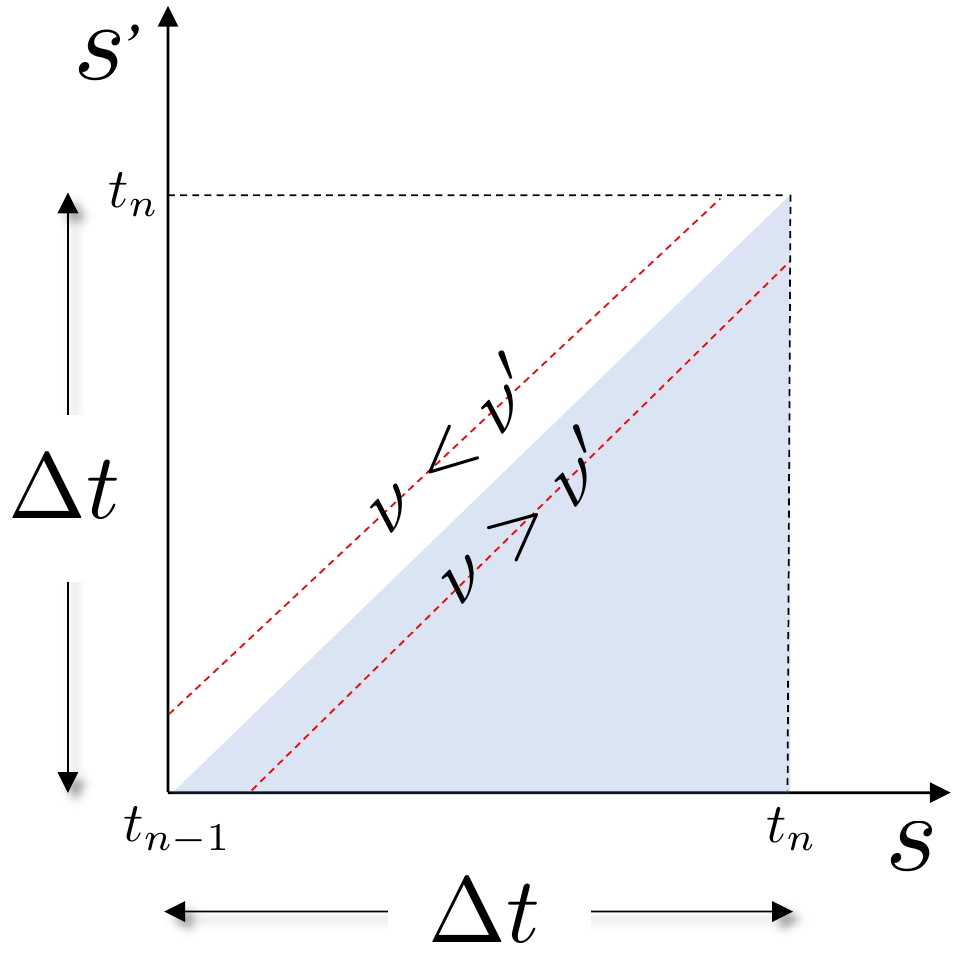}
		\caption{Calculation of double integrals appearing in the vacuum term \eqref{H1vac}. The shaded region (triangle) represents the domain of integration. The integrand $\de(s'-s+(\tau_\nu-\tau_{\nu'}))$ vanishes everywhere except on the red line $s'=s-(\tau_\nu-\tau_{\nu'})$. This line lies within the triangular domain for $\nu>\nu'$ and outside of it for $\nu<\nu'$. Thereby, the integral is equal to $\Delta t$ in the former case and vanishes in the latter.}\label{square}
	\end{center}
\end{figure}

Thus we are left only with the vacuum contribution $\HH_{\rm vac}^{(1)}$. To work this out, we first note that the each double integral in \eq\eqref{H1vac} runs over the shaded triangle sketched in \fig\ref{square}. For a given pair ($\nu$, $\nu'$), the two-variable $\delta$ function 
\begin{equation}
\delta(s'-\tau_{\nu'}-s+\tau_{\nu})=[ {\hat b}_{s'-\tau_{\nu'}}{,}{\hat b}^\dag_{s-\tau_{\nu}}]\label{delta}
\end{equation}
is peaked on the line $s'=s-(\tau_\nu-\tau_{\nu'})$.
As shown in \fig\ref{square}, this line falls within the triangle for $\nu>\nu'$ and outside of it for $\nu<\nu'$  (since $\tau_\nu-\tau_{\nu'}>0$ for $\nu>\nu'$). Hence, only terms $\nu>\nu'$ contribute to $\HH_{\rm vac}^{(1)}$ and we conclude that $\HH_{\rm vac}^{(1)}\equiv\hat  H_{\rm vac}$ [\cf\eq\eqref{Hvac}].
\\
\\
\indent The above shows that, for delays negligible with respect to $\Delta t$ this being in turn much shorter than the interaction characteristic time scale $\gamma^{-1}$, in \eq\eqref{Un-app} we can approximate $\HH^{(0)}_n\simeq \hat V_n$ and $\HH_n^{(1)}\simeq \hat H_{\rm vac}$. Thereby,
\begin{align}
\hat U_n\simeq \openone -i \,(\hat H_{\rm vac}+\hat V_n)\, \Delta t -\tfrac{1}{2} \hat V_n^2 \,\Delta t^2\,,\label{Un-2}
\end{align}
showing that in this regime the joint emitter-field dynamics can be effectively pictured as a sequence of short pairwise interactions (collisions) of duration $\Delta t$ (collision time), as sketched in \figs\ref{sketch} and \ref{delays}(a). In each interaction the emitters collectively couple to a fresh time bin (only one) according to the coupling Hamiltonian $\hat V_n$ and at the same time coherently interact with one another through the second-order many-body Hamiltonian $\hat H_{\rm vac}$. 
Note that time bins are uncoupled from one another and that each collides with the emitter only once in a ``conveyor-belt'' fashion (see \fig\ref{sketch}).

As said, to arrive at \eq\eqref{Un-2}, all time delays $\tau_{\nu}-\tau_{\nu'}$ were neglected. We point out that this is {different} from setting $\tau_{\nu}-\tau_{\nu'}=0$. Instead, it corresponds to performing the limit $\tau_{\nu}-\tau_{\nu'}\rightarrow 0^+$ for all pairs $(\nu,\,\nu')$ with $\nu>\nu'$. Indeed, it is easily checked that setting $\tau_{\nu}-\tau_{\nu'}=0$ entails $\HH_{\rm vac}^{(1)}=0$ since in this case both terms $\nu>\nu'$ and $\nu<\nu'$ must be accounted for but exactly cancel out (the two dashed lines in \fig\ref{square} now both reduce to $s'=s$). Physically, this means that the effective Hamiltonian $\hat H_{\rm vac}$ stems from the fact that, while traveling from left to right [see \fig\ref{delays}(a)], the $n$th time bin interacts {\it first} with coupling point $\nu$ and only {\it afterwards} with $\nu+1$, no matter how short the delay $\tau_{\nu+1}-\tau_{\nu}$.
This is in line with similar observations made in derivations of cascaded MEs through other methods (see \eg \cite{GardinerPRLcascaded}). Interestingly, the collisional picture allows for a complementary interpretation of this phenomenon in terms of far-detuned time-bin modes $\hat b_{n,k}$, which we introduce next.

\subsection{Time-bin modes $\hat b_{n,k}$}\label{sec-bnk}

It should be clear from their definition \eqref{rn0} that, for a finite $\Delta t$, modes $\hat b_n$ generally capture only part of the field degrees of freedom. Formally, this can be seen by expanding the continuous time modes as \cite{grossQubit2018}
\begin{equation}
\hat b_t=\tfrac{1}{\sqrt{\Delta t}}\sum_n\sum_{k=-\infty}^\infty \Theta_n(t) e^{-i {2\pi k t}/{\Delta t}}\,\hat b_{n,k}\,,\label{fourier}
\end{equation}
with $\Theta_n(t)=1$ for $t\in [t_{n-1},t_n]$ and 0 otherwise, and where 
\begin{equation}
\label{rnk}
\hat b_{n,k}=\tfrac{1}{\sqrt{\Delta t}}\int_{t_{n-1}}^{t_n}\!dt\, e^{i {2\pi k t}/{\Delta t}}\,b_t\,.
\end{equation}
Ladder operators $\hat b_{n,k}$ fulfill $[\hat b_{n,k},\hat b_{n',k'}^\dag]=\delta_{n,n'}\delta_{k,k'}$, $[\hat b_{n,k},\hat b_{n',k'}]=[\hat b^\dag_{n,k},\hat b_{n',k'}^\dag]=0$. 
Moreover, for $k=0$ we retrieve modes $\hat b_n$  [\cf\eq\eqref{rn0}], i.e., $\hat b_{n,0}\equiv \hat b_n$. A straightforward Fourier analysis shows that time-bin modes $\hat b_{n,k\neq 0}$ are dominated by field normal modes whose detunings from the emitter grow as $\sim |k|/\Delta t$, (while modes $\hat b_{n,0}$ contain field frequencies quasi-resonant with the emitter) \cite{grossQubit2018}. For $\Delta t\rightarrow 0$ [still fulfilling \eqref{hie}], corresponding to the continuous-time limit of the dynamics, these frequencies become divergent. Accordingly, it is reasonable to assume there are no photons populating modes $\hat b_{n,k\neq 0}$. This is equivalent to stating that the most general field state is of the form
\begin{equation}
\rho_f = \eta_{\rm bins} \bigotimes_{n,k\neq 0}\ket{0}_{n,k}\!\bra{0}\label{field-state}
\end{equation}
with $\eta_{\rm bins}$ the (generally mixed) state of modes $\hat b_n\equiv \hat b_{n,0}$ and $\ket{0}_{n,k}$ the vacuum state of mode $\hat b_{n,k}$. 

\subsection{Differences with the single-coupling-point case}\label{sec-diff}

For a {\it single} coupling point (${\cal N}=1$) $\hat H_{\rm vac}$ of course does not arise and we are only left with $\hat V_n$ (containing only $\hat b_n\equiv \hat b_{n,0}$), meaning that the coupling to time-bin modes $k\neq 0$ is negligible. Yet, for two or more coupling points (${\cal N}\ge 2$), these off-resonant modes yield non-negligible effects despite they do not explicitly appear in $\hat H_{\rm vac}$ (not even in $\hat V_n$, of course). Indeed, they are in fact responsible for the emergence of $\hat H_{\rm vac}$. This can be seen from \eq\eqref{H1vac} featuring a singularity in the integrand function due to the field commutator. Such a singular behavior forbids to retaining only $k=0$ terms in expansion \eqref{fourier} no matter how small $\Delta t$ (indeed it is easily checked that expanding each field operator entering \eq\eqref{H1vac} and retaining only modes $\hat b_{n,0}=\hat b_n$ would yield a vanishing ${\hat H}_{\rm vac}$). 

Thus all time-bin modes $\hat b_{n,k}$ in fact contribute to the dynamics for ${\cal N}>1$. However, unlike $k=0$ modes, off-resonant modes $k\neq 0$ are only virtually excited, explaining why they do not explicitly appear in $\hat H_{\rm vac}$. 

\subsection{Non-negligible time delays}\label{non-negli}

A comprehensive treatment of the regime of non-negligible delays is beyond the scope of the present paper. Yet, we wish to highlight a major difference from the negligible delays regime, this being that at each time step the emitters collide with as many time bins as the number of coupling points (instead of only one). To illustrate this, we work out next $\HH_n^{(0)}$ [\cf\eq\eqref{Mag1} and its equivalent expression \eqref{Mag1-1}].

In contrast with the negligible delays regime, now one can take a time step negligible compared with all the system's time delay, i.e., $\Delta t \ll \tau_\nu-\tau_{\nu-1}$ for all $\nu$ (note that this is compatible with condition $\Delta t\ll \gamma^{-1}$ that we assume throughout). For sufficiently short $\Delta t$, the coupling points coordinates can be discretized as  $\tau_\nu=m_\nu\Delta t$, where $\{m_\nu\}$ are ${\cal N}$ integers such that $m_1<m_2<...<m_{\cal N}$, and set $\tau_1=m_1=0$.
Accordingly, \eqref{Mag1-1} becomes [recall that $t_n=n\Delta t$]
\begin{equation}
\HH_n^{(0)}= \sqrt{\tfrac{\gamma}{\Delta t}}\,\sum_\nu\,(\A_{\nu}\,\hat b_{n-m_\nu}^\dag+{\rm H.c.})\,,\label{Mag1-dis}
\end{equation}
showing that, during a given time interval $[t_{n-1},t_n]$, each coupling point $\nu$ interacts with a different time bin $n-m_{\nu}$ [see \fig\ref{delays}(b)].

In the presence of giant emitters (even a single one), this dynamics is tough to tackle analytically. Through an elegant diagrammatic technique, Grimsmo found an analytical solution for the open dynamics of a driven giant atom with two coupling points \cite{GrimsmoPRL15}, while Pichler and Zoller found an efficient matrix-product-state approach which they applied to a pair of driven normal atoms coupled to a bidirectional field \cite{pichlerPhotonic2016} (the collisional picture for a bidirectional field is addressed in Section \ref{CM-bi}). A major reason behind the complexity of this dynamics lies in its generally non-Markovian nature (conditions for Markovian behaviour are discussed in Section \ref{section-ME}).

\section{Master equation \\for negligible time delays}\label{section-ME}

In section \ref{negli}, we focused on the total {propagator} showing that for negligible time delays it can be decomposed as a sequence of collisions between the emitters (jointly) and a field time bin, each described by the two-body elementary unitary $\hat U_n$ in \eq\eqref{Un-2}, which is fully specified by ${\hat H}_{\rm vac}$ and $\hat V_n$. In this section, we derive master equations for the emitters and time bin in the regime of negligible time delays. 

\subsection{Conditions for Markovian dynamics}

Based on \eqref{field-state} and related discussion, from now on time-bin modes $\hat b_{n,k\neq 0}$ will be ignored.
The joint state of the emitters and {\it all} time bins (modes $\hat b_n\equiv \hat b_{n,0}$) evolves at each time step as $\sigma_n=\hat U_n \sigma_{n-1}\hat U_n^\dag$ with $\sigma_n=\sigma_{t=t_n}$. A corresponding finite-difference equation of motion is worked out by replacing $\hat U_n$ with \eqref{Un-2} and retaining only terms up to second order in $\Delta t$
\begin{eqnarray}
\frac{\Delta \sigma_n}{\Delta t}=&&-i \,[\hat H_{\rm vac}+\hat V_n,\sigma_{n-1}]\nonumber\\
&&+\Delta t\!\left(\hat V_n\sigma_{n-1}\hat V_n-\tfrac{1}{2}\left[\hat V_n^2,\sigma_{n-1}\right]_+\right),\label{dsigma-2}
\end{eqnarray}
where $\Delta \sigma_n=\sigma_{n}-\sigma_{n-1}$ (recall that $\hat V_n\sim 1/\sqrt{\Delta t}$). Under the usual assumption of zero initial correlations between the emitters and the field, the initial condition reads $\sigma_0=\rho_0\otimes\eta_{\rm bins}$, where $\rho_0$ and $\eta_{\rm bins}$ are the initial states of all emitters and all time bins, respectively. 

We next ask whether or not the reduced dynamics of the emitters $\rho_n={\rm Tr}_{\rm bins}\{\sigma_n\}$ is Markovian and describable by a Lindblad master equation. We note that this is generally not the case when time bins are initially correlated, namely $\eta_{\rm bins}$ is {\it not} a product state, since in these conditions the emitters can get correlated with a time bin even before colliding with it \cite{filippovDivisibility2017,ciccarelloCollision2017}. This indeed rules out that that the evolution of the emitters (open system) at each elementary collision be described by a completely positive and trace preserving (CPT) quantum map \cite{breuerTheory2007}, which is the key requirement in order for a Lindblad master equation to hold. A typical instance is a single-photon wavepacket of bandwidth comparable with $\gamma$ \cite{DabrowskaPRA2017,Bouten2019,DabrowskaJPA2019,DabrowskaJOSA2020,FangNJP18}. 

We thus consider the case that time bins are initially uncorrelated, that is
\begin{equation}
\eta_{\rm bins}=\bigotimes_n \eta_n\,\label{uncorr}
\end{equation}
with $\eta_n$ the reduced state of the $n$th time bin mode having ladder operator $\hat b_n=\hat b_{n.0}$. This entails
\begin{equation}
\rho_n={\rm Tr}_{\rm bins}\left\{\hat U_n \sigma_{n-1}\hat U_n^\dag\right\}={\rm Tr}_{\rm n}\left\{\hat U_n \rho_{n-1}\eta_n\hat U_n^\dag\right\}\,,\!\!\label{rhon}
\end{equation}
where ${\rm Tr_n}$ is the partial trace over the time bin $n$ (mode $\hat b_n\equiv \hat b_{n,0}$). This defines a CPT map describing how the emitters' state $\rho_{n}$ is changed by the $n$th collision. Likewise, the $n$th time bin evolves according to
\begin{equation}
\eta'_n={\rm Tr}_{S}\left\{\hat U_n \rho_{n-1}\eta_n\hat U_n^\dag\right\}\,\label{varrho}
\end{equation}
with ${\rm Tr}_S$ the partial trace over the emitters. This is a CPT map describing the change of the single time bin state due to collision with the emitters (after the collision this state will no longer change since time bins are non-interacting). Note that map \eqref{varrho} depends parametrically on the current reduced state of emitters (updated at each collision).

\subsection{Master equation for the emitters}

To work out the Lindblad master equation of the emitters corresponding to map \eqref{rhon} we simply trace off all time bins from \eq\eqref{dsigma-2}, which yields
\begin{align}
\frac{\Delta \rho_n}{\Delta t}=&-i [\hat H_{\rm vac}+\langle \hat V_n\rangle,\rho_{n-1}]+{\cal D}[\rho_{n-1}]\label{d-ME-2}\,
\end{align}
with $\langle...\rangle={\rm Tr}_n\left\{...\,\eta_n\right\}$, $\Delta \rho_n=\rho_{n}-\rho_{n-1}$ and
\begin{equation}
{\cal D}[\rho_{n-1}]=\Delta t {\rm Tr}_n\!\left\{ \hat V_n \rho_{n-1} \eta_{n}\hat V_n-\tfrac{1}{2}\left[\hat V_n^2, \rho_{n-1} \eta_{n}\right]_+\right\}. 
\end{equation}
Although not explicit, this equation is in Lindblad form as is easily checked by spectrally decomposing $\eta_n$ \cite{note-lind}. The Linbdlad form is a guaranteed by the fact that the emitters evolution at each collision is described by a CPT map [last identity in \eq\eqref{rhon}]. 

Using \eqref{Vn} the first-order Hamiltonian and second-order dissipator can be put in the more explicit form
\begin{eqnarray}
\langle \hat V_n\rangle&=& \sqrt{\tfrac{\gamma}{\Delta t}}\left(\langle{{\hat b}_n}\rangle\,  \A^\dag+{\rm H.c.}\right)\label{Vn3}\\
{\cal D}[\rho_{n-1}]&=&\!\gamma\sum_{\mu\mu'}\langle {\hat c}_\mu {\hat c}_{\mu'}\rangle\!\left(\!\hat {\cal C}_{\mu'}\rho_{n-1}\hat {\cal C}_\mu{-}\tfrac{1}{2}\left[\hat {\cal C}_\mu\hat {\cal C}_{\mu'},\rho_{n-1}\right]_+\!\right)\nonumber\\
\label{Drho}\,.
\end{eqnarray}
with $\mu, \mu'=1,2$ and where we set 
\begin{equation}
\hat c_1=\hat b_n\,,\,\, \hat c_2=\hat b_n^\dag\,,\,\,\hat {\cal C}_1=\A^\dag\,,\,\,\hat {\cal C}_2=\A\,. \label{cC}
\end{equation}
Now \eq\eqref{d-ME-2} is expressed fully in terms of the time-bin moments $\langle \hat b_n \rangle$, $\langle \hat b_n^\dag\hat b_{n}\rangle$ and $\langle \hat b_n^2\rangle$, which depend on $\eta_n$ in turn dependent on the initial field state [\cf\eq\eqref{uncorr}]. 

The time-bin moments can be determined for the most general white-noise Gaussian state of the field. As anticipated in Section \ref{summary}, such a state is fully specified by the 1st and 2nd moments \eqref{dB} \cite{WisemanMilburnBook}.
Noting that $\hat b_n=\int_{t_{n-1}}^{t_n}\!d b_t/\sqrt{\Delta t}$, it is evident that for such a field state, $\langle \hat b_{n}^\dag\hat b_{n'}\rangle=\langle \hat b_{n}\hat b_{n'}\rangle=0$ for $n\neq n'$. This, because of the Gaussianity hypothesis, is equivalent to \eq\eqref{uncorr}. Thus time bins are initially uncorrelated. Their 1st and 2nd moments are given by \eqref{dbn}, 
where the rigorous $\alpha_n$'s definition is $\alpha_n=\int_{t_{n-1}}^{t_n}\!dt \,\alpha_{t}/\Delta t$  (for $\Delta t$ short enough, this reduces to $\alpha_n\simeq \alpha_{t_n}$). Note that $\langle \hat b_n \rangle\propto \sqrt{\Delta t}$, which cancels the $1/\sqrt{\Delta t}$ factor in \eq\eqref{Vn3}.

Plugging moments \eqref{dbn} into the finite-difference \eq\eqref{d-ME-2} and taking the continuous-time limit such that $\gamma\Delta t\rightarrow 0$, $t_n\rightarrow t$, $\rho_{n-1}\rightarrow \rho_t$, $\Delta \rho_n/\Delta t\rightarrow d\rho/dt$ we end up with the general master equation \eqref{g-ME}.

\subsection*{Time bin master equation}

An equation for the rate of change of the single time bin state, $\Delta \eta_n/\Delta t$ with $\Delta \eta_n=\eta'_{n}-\eta_{n}$, can be similarly worked out. We again start from \eq\eqref{dsigma-2} but now trace over all emitters and all time bins $n'\neq n$, obtaining
\begin{align}
\frac{\Delta \eta_n}{\Delta t}=&-i [\langle \hat V_n\rangle_{\rho},\eta_{n}]+{\cal D}_{\rho}[\eta_{n}]\label{d-tb}\,
\end{align}
with $\langle...\rangle_{\rho}={\rm Tr}_S\left\{...\,\rho\right\}$ and
\begin{equation}
{\cal D}_{\rho}[\eta_{n}]=\Delta t\,{\rm Tr}_S\left\{ \!\hat V_n \rho_{n-1} \eta_{n}\hat V_n{-}\tfrac{1}{2}\left[\hat V_n^2, \rho_{n-1} \eta_{n}\right]_+\right\} 
\end{equation}
where ${\rm Tr}_S\left\{...\right\}$ is the partial trace over the system. Note that this equation parametrically depends on the state of the emitters, $\rho_{n-1}$, which changes at each time step. \eq\eqref{d-tb} expresses map \eqref{varrho} in the short-collision-time limit.

\section{Collision model for a bidirectional field}\label{CM-bi}

For a bidirectional field (see Section \ref{sec-bi}), unitaries $\U_t$ and $\hat U_n$ are formally the same as \eqref{Ut2} and \eqref{Magnus}, respectively, but $\hat V_t$ is now given by \eq\eqref{Vt-rl}. 
The $\hat U_n$'s lowest-order expansion \eqref{Un-app} is formally unchanged. Through a reasoning analogous to that in Section \ref{derivation}, in light of \eqref{Vt-rl}, \eqs\eqref{Mag1-1} and \eqref{HH1} are generalized as
\begin{eqnarray}
\HH^{(0)}_n=\!&& \sqrt{\tfrac{\gamma}{\Delta t}}\,\sum_\nu\A_{\nu}\left(\tfrac{1}{\sqrt \Delta t}\!\int_{t_{n-1}}^{t_n} \!ds\,\,  \hat b^\dag_{s-\tau_\nu}\right)\nonumber\\
&&+ \!\sqrt{\tfrac{\gamma'}{\Delta t}}\,\sum_\nu\A'_{\nu}\left(\tfrac{1}{\sqrt \Delta t}\!\int_{t_{n-1}}^{t_n} \!ds\,\,  {\hat {{b}'}}^\dag_{s+\tau_\nu}\right)+{\rm H.c.}\,,\,\,\,\label{HH0-rl}\\
\HH_n^{(1)}=&&\HH_{\rm sq}^{(1)}+\HH_{\rm th}^{(1)}+\HH_{\rm vac}^{(1)} \label{HH1-rl}
\end{eqnarray}
with
\begin{widetext}
\begin{eqnarray}
\HH_{\rm vac}^{(1)}=&&\,\tfrac{i}{2\Delta t}\!\sum_{\nu,\nu'}\A_{\nu'}^\dag \A_{\nu}\!\int_{t_{n-1}}^{t_n} \!\!ds\! \int_{t_{n-1}}^{s}\!\! \!ds'
 \left(
 \gamma\, [ {\hat b}_{s'-\tau_{\nu'}}{,}{\hat b}^\dag_{s-\tau_{\nu}}] + 
 \gamma'\,[ {{\hat {{b}'}}}_{s'+\tau_{\nu'}}{,}{{\hat {{b}'}}}^\dag_{s+\tau_{\nu}}] - {\rm H.c.} \right)\,, \label{HHbivac}
\\
\HH_{\rm th}^{(1)}=&&\,\tfrac{i}{2\Delta t}\!\sum_{\nu\nu'}\!\int_{t_{n-1}}^{t_n} \!\!ds\! \int_{t_{n-1}}^{s}\!\! \!ds'  \left( \gamma\,[\A_{\nu'}^\dag,\A_{\nu}] \,{\hat b}^\dag_{s-\tau_{\nu}}{\hat b}_{s'-\tau_{\nu'}}
+ \gamma' [\A_{\nu'}^{'\dag},\A_{\nu}^{'}] \,{{\hat {{b}'}}}^\dag_{s+\tau_{\nu}}{{\hat {{b}'}}}_{s'+\tau_{\nu'}} -{\rm H.c.} \right) \nonumber
\\&&
+\tfrac{i}{2\D}\,\sqrt{\gamma\gamma' }
\sum_{\nu\nu'}\!\int_{t_{n-1}}^{t_n} \!\!ds\! \int_{t_{n-1}}^{s}\!\! \!ds' 
\left([\mathcal{A}_{\nu}^\dag ,\mathcal{A}_{\nu'}^{'} ]\, \hat {b}_{s-\tau_{\nu}}\,\hat {{b}'}^\dag_{s'+\tau_{\nu'}} 
+
[\mathcal{A}_\nu^{'\dag} , \mathcal{A}_{\nu'} ] \,\hat {b}^\dag_{s'-\tau_{\nu'}} \,\hat {{b}'}_{s+\tau_{\nu}} - {\rm H.c.} \right)\,, \label{HHbith}
\\
\HH_{\rm sq}^{(1)}=&&\,\tfrac{i}{2\D}
\!\sum_{\nu\nu'}\int_{t_{n-1}}^{t_n} \!\!ds\! \int_{t_{n-1}}^{s}\!\! \!ds' \left( \gamma\, [\A_{\nu'},\A_{\nu}] \,{\hat b}^\dag_{s-\tau_\nu}{\hat b}^\dag_{s'-\tau_{\nu'}} +\gamma' \,[\A^{'}_{\nu'},\A^{'}_{\nu}]\, {{\hat {{b}'}}}^\dag_{s+\tau_\nu}{{\hat {{b}'}}}^\dag_{s'+\tau_{\nu'}} - {\rm H.c.} \right)\nonumber
\\&&
+\tfrac{i}{2 \D}\,\sqrt{\gamma\gamma' }
\sum_{\nu\nu'}\!\int_{t_{n-1}}^{t_n} \!\!ds\! \int_{t_{n-1}}^{s}\!\! \!ds'
\left([\mathcal{A}_{\nu'}^{'\dag},\mathcal{A}_{\nu}^\dag  ] \,\hat {b}_{s-\tau_{\nu}} \hat {{b}'}_{s'+\tau_{\nu'}} 
+
[\mathcal{A}_{\nu'}^\dag, \mathcal{A}_\nu^{'\dag}  ]\, \hat {b}_{s'-\tau_{\nu'}} \hat {{b}'}_{s+\tau_{\nu}} - {\rm H.c.} \right)\,. \label{HHbisq}
\end{eqnarray}
\end{widetext}

\subsection{Negligible time delays}

Regarding $\HH^{(0)}_n$, an argument analogous to that leading to \eqref{HH0-Vn} now yields that in the present regime $\HH^{(0)}_n\simeq \hat V_n$ with $\hat V_n$ given by \eq\eqref{Vn-rl}.
Regarding $\HH^{(1)}_n$, as in the unidirectional case terms $\HH_{\rm th}^{(1)}$ and $\HH_{\rm sq}^{(1)}$ are again negligible in the limit of vanishing delays (see Appendix \ref{app1}). Compared to the unidirectional case [\cf\eq\eqref{H1vac}], $\HH_{\rm vac}^{(1)}$ has an extra term, due to the left-going modes, featuring the $\delta$ function $[ {{\hat {{b}'}}}_{s'+\tau_{\nu'}}{,}{{\hat {{b}'}}}^\dag_{s+\tau_{\nu}}]$. This peaks on the line $s'=s-(\tau_{\nu'}-\tau_{\nu})$, which differs from the $\delta$ function coming from right-going modes [\cf\eq\eqref{delta}] for the exchange $\nu\leftrightarrow \nu'$. Accordingly, in \fig\ref{square}, the lines corresponding to $\nu<\nu'$ and $\nu>\nu'$ are swapped, hence now only terms $\nu<\nu'$ (instead of $\nu>\nu'$) contribute to $\HH_{\rm vac}^{(1)}$. Thus we end up with $\HH_{\rm vac}^{(1)}\equiv\hat  H_{\rm vac}$ with $\hat  H_{\rm vac}$ given by \eq\eqref{Hvac-rl}.

Thereby, for $\tau_{\cal N}-\tau_1\ll\Delta t\ll (1/\gamma,1/\gamma')$, the joint dynamics can be be represented by an effective collision model (see \fig\ref{bipartite}), where at each collision the emitters jointly collide with a right-going and a left-going time bin, at once being subject to an internal coherent dynamics governed by the second-order Hamiltonian \eqref{Hvac-rl}. Note that, formally, this can still be thought as a collision model featuring a single stream of time bins [like \fig\ref{sketch}] provided that one defines a two-mode time bin ($\hat b_n$, ${\hat {{b}'}}\!_n$).

\subsection{Non-negligible time delays}

An argument analogous to that used in Section \ref{non-negli} generalizes \eq\eqref{Mag1-dis} as
\begin{equation}
\HH_n^{(0)}=\tfrac{1}{ \sqrt{\Delta t}}\sum_\nu\left(\sqrt{\gamma}\,\A_{\nu}\,\hat b_{n-m_\nu}^\dag{+}\sqrt{\gamma'}\Ap_{\nu}\,{\hat {{b}'}}_{n+m_\nu}{+}{\rm H.c.}\right).\label{H0-rl}
\end{equation} 
Here, ladder operators ${\hat {{b}'}}\!_n$ [\cf\eq\eqref{rtn}] define a discrete collection of left-going bosonic modes analogous to $\hat b_n$ (the former commuting with the latter).
\begin{figure}[t!]
	\begin{center}
		\includegraphics[width=0.42\textwidth]{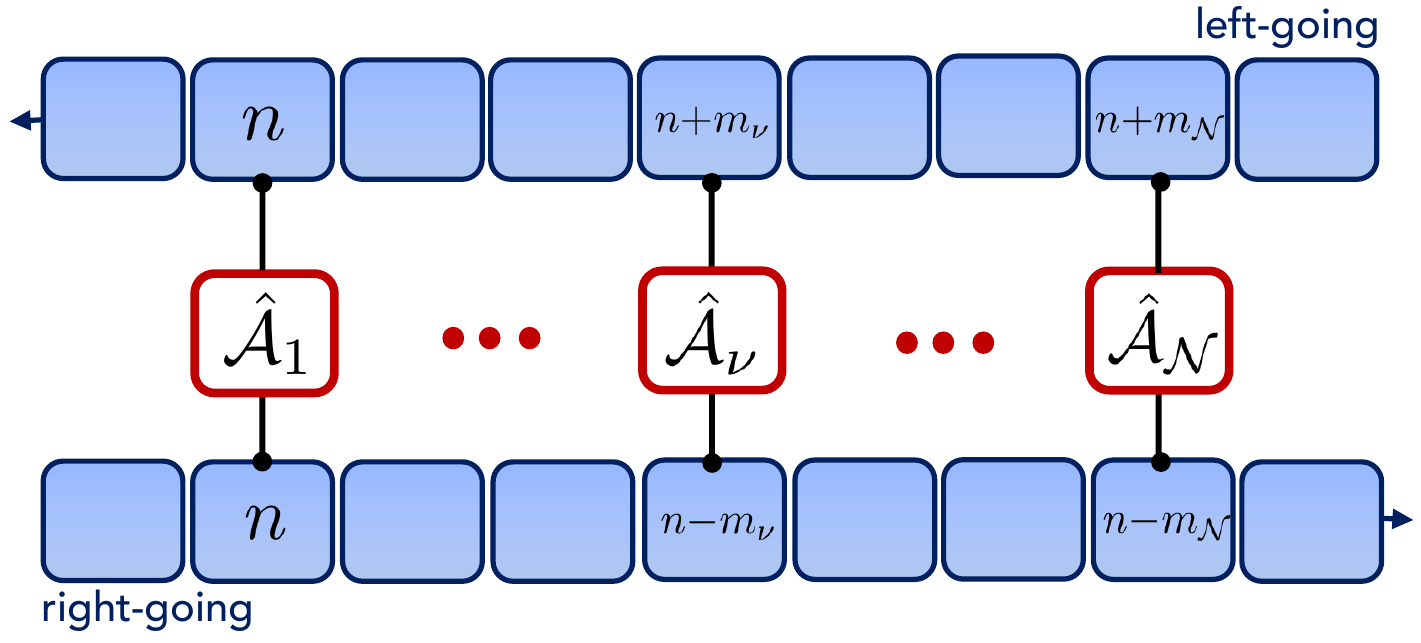}
		\caption{Bidirectional field for non-negligible time delays. Left-going time bins (top) and right-going time bins (bottom). We set $\tau_1=m_1=0$.}\label{bi-long}
	\end{center}
\end{figure}
Note the different subscripts in $\hat b_{n-m_\nu}$ and ${\hat {{b}'}}_{n+m_\nu}$, reflecting that right- and left-going time bins travel in opposite directions as sketched in \fig\ref{bi-long}.

Similarly to the unidirectional case discussed in Section \ref{non-negli}, analytical descriptions of this dynamics are demanding \cite{pichlerPhotonic2016,Pichler2015,guimondDelayed2017}.

\section{Master equation for a bidirectional field}\label{section-ME-bi}

With the extended definitions of $\hat  H_{\rm vac}$ and $\hat V_n$ for a bidirectional field discussed in the previous section (regime of negligible time delays), the finite-difference equation of motion \eqref{dsigma-2} for the joint dynamics still holds. The initial state of the time bins $\eta_{\rm bins}$ is obtained from the initial field state by using \eqref{fourier} and tracing off time-bin modes $k\neq 0$ (with left-going time-bin modes $\hat b'_{n,k}$ also accounted for).

Likewise, a Markovian open dynamics will arise in the case of field states which in the collisional picture turn into uncorrelated states of the time bins 
\begin{equation}
\eta_{\rm bins}=\bigotimes_{n}(\eta_{r,n} \otimes\eta_{l,n})\,\label{uncorr-2}
\end{equation}
with $\eta_{r,n}$ ($\eta_{l,n}$) the reduced state of the $n$th right-going (left-going) time bin. Preparing such states, the emitters evolve at each collision according to a CPT map [\cf\eq\eqref{rhon}] and so do time bins [see \eq\eqref{varrho}].

The finite-difference master equation of the emitters \eqref{d-ME-2} holds, where $\langle \hat V_n\rangle$ and ${\cal D}[\rho_{n-1}]$ are now given by
\begin{eqnarray}
&&\langle \hat V_n\rangle=\sqrt{\tfrac{1}{\Delta t}}\left(\sqrt{\gamma}\,\langle{{\hat b}_n}\rangle\,  \A^\dag+\sqrt{\gamma'}\,\langle{{{\hat {{b}'}}}_n}\rangle\,  \Ap^\dag+{\rm H.c.}\right),\,\,\,\,\,\,\,\\
&&{\cal D}[\rho_{n-1}]={\cal D}_r[\rho_{n-1}]+{\cal D}_l[\rho_{n-1}]\,
\end{eqnarray}
with ${\cal D}_r[\dots]$ the same as \eqref{Drho} and ${\cal D}_l[\dots]$ obtained from \eqref{Drho} through the replacements $\gamma\rightarrow \gamma'$, $\hat b_n\rightarrow {\hat {{b}'}}\!_n$, $\A\rightarrow \Ap$.
The master equation is expressed in terms of first and second moments of right-going and left-going time bins, respectively depending on $\eta_{r,n}$ and $\eta_{l,n}$ [\cf\eq\eqref{uncorr-2}].

The most general white-noise Gaussian state of the field is now specified by right-going moments \eqref{dB} plus the analogously defined left-going moments $\alpha'_t$, $N'$ and $M'$. The latter determine the time-bin moments $\langle {\hat {{b}'}}\!_n \rangle=\alpha'_n \sqrt{\Delta t}$, $\langle {\hat {{b}'}}_n^\dag {\hat {{b}'}}_{n}\rangle=N'$ and $\langle {\hat {{b}'}}_n^2\rangle=M'$. 
Plugging these into the finite-difference \eq\eqref{d-ME-2} and taking next the continuous-time limit as done in the unidirectional case, we end up with master equation \eqref{g-ME-rl}.

\section{Photodetection and quantum trajectories}\label{sec-jump}

As anticipated in the Introduction, a major advantage of collision models is that they naturally accommodate quantum weak measurements \cite{brunSimple2002}, which in the present  quantum-optics framework correspond to photodetection \cite{Carmichael93,WisemanMilburnBook}.

Let $\{\ket{k}_n\}$ be an orthonormal basis of the $n$th time bin (henceforth we will mostly prefer the compact notation $\ket{k_n}$). In the collisional picture, photodetection consists in measuring each time bin right after its collision with $S$. The photodetection scheme is defined by the measurement basis $\{\ket{k}\}$.
Assuming an uncorrelated initial state of the time bins [\cf\eq\eqref{uncorr}], the (unnormalized) evolved state of the joint system after a specific sequence of measurement outcomes $\{k_1,...,k_n\}$ is given by 
\begin{widetext}
	\begin{equation}
	{\tilde\sigma}_n=\ket{k_n}\!\bra{k_n}\hat U_n \cdots \ket{k_1}\!\bra{k_1}\hat U_1\left(\rho_0 \bigotimes_{m}\eta_{m}\right) \hat U_1^\dag \ket{k_1}\!\bra{k_1}\cdots \hat U_n^\dag \ket{k_n}\!\bra{k_n} \label{tildesig}
	\end{equation}
\end{widetext}
with $\hat U_n$ given by \eq\eqref{Un-2}. The probability $p_{k_1\cdots \,k_n}$ of getting this sequence of measurement outcomes is the norm of $\tilde\sigma_n$,
\begin{equation}
p_{k_1\cdots \,k_n}={\rm Tr}\left\{{\tilde\sigma}_n\right\}\,,\label{prob}
\end{equation}
hence the normalized state is $\sigma_n=\tilde{\sigma}_n/p_{k_1\cdots \,k_n}$.

Let each time bin be initially in a pure state $\eta_{m}=\ket{\chi_{m}}\!\bra{\chi_{m}}$ (the mixed case is commented later). Plugging this into \eqref{tildesig} and tracing off all the time bins yields \eqref{cond-sum}, i.e., the unnormalized state of the emitters $S$ at step $n$ (while time bins are of course in state $\ket{k_1}\cdots\ket{k_n}$, uncorrelated with $S$). Kraus operators $\hat {K}_{k_m}$ [\cf\eqref{kraus-sum}] are reported here again for convenience
\begin{align}
\hat {K}_{k_m}=\bra{k_m}\!\hat{ U}_m\ket{\chi_m}\,.\label{Kraus}
\end{align}
Thus, since ${\rm Tr}_n\{\tilde{\sigma}_n\}={\rm Tr}_S\{\tilde{\rho}_n\}$, $p_{k_1\cdots \,k_n}$ can be expressed as [\cf\eq\eqref{prob}]
\begin{equation}
p_{k_1\cdots \,k_n}={\rm Tr}_S\left\{\hat {K}_{k_n}\cdots\hat {K}_{k_1}\,\rho_0\, \hat {K}_{k_1}^\dag\cdots \hat {K}_{k_n}^\dag\right\}\,.
\end{equation}
As anticipated in Section \ref{sec-jump-summ}, time-bin measurement with outcome $k$ (at the end of the $n$th collision) thus projects the emitters into the (unnormalized) state
\begin{equation}
\tilde\rho_n=\hat K_k \,\rho_{n-1} \,\hat K_k^\dag\,\label{proj}
\end{equation}
with probability
\begin{align}
p_k=  \Tr_S\{\tilde\rho_n\}\ = \Tr_S\{\hat{K}_k^\dag \hat{K}_k \,\rho_{n-1}\}\,.
\label{pk}
\end{align}

($\rho_{n-1}$ denotes as usual the normalized state of $S$ right before collision with time bin $n$). This map defines the conditional dynamics. Summing next over all possible $k$'s yields $\rho_n=\sum_k \hat K_k \,\rho_{n-1} \,\hat K_k^\dag $, this CPT map defining the unconditional open dynamics.

We derive next the lowest-order expansion of each Kraus operator $\hat {K}_{k}$.
Let us first arrange \eqref{Un-2} in the form 
\begin{align}
\hat U_n\simeq \openone &-i \sqrt{\gamma}\left(\A \,\hat b_n^\dag+\A^\dag \,\hat b_n \right) \sqrt{\Delta t} \nonumber\\
&-i\left(\hat H_{\rm vac}-i\tfrac{\gamma}{2} (\A \,\hat b_n^\dag+\A^\dag \,\hat b_n)^2 \right)\Delta t\,,\label{Un-3}
\end{align}
where we used \eqref{Vn}. Moreover, we allow the time-bin state $\ket{\chi_n}$ to generally depend on $\sqrt{\Delta t}$ (this is for instance the case of coherent states as illustrated later). Hence, to lowest order, expansion \eqref{chi-exp-sum} holds.
Note that the zeroth-order term was set equal to the time-bin vacuum state to ensure that the field energy density $\langle \hat b_n^\dag \hat b_n\rangle/\Delta t$ does not diverge in the limit $\Delta t\rightarrow 0$, which would lead to nonsensical photon-counting evolution~\cite{WisemanMilburnBook,grossQubit2018} [we come back to this issue shortly after \eq\eqref{tilderho}]. Also, $|\chi_n^{(1)}\rangle$ and $|\chi_n^{(2)}\rangle$ are subject to the constraints
\begin{align}
&{\rm Re}\,\langle 0_n|\chi_n^{(1)}\rangle=0
\\
&\langle \chi_n^{(1)}|\chi_n^{(1)}
\rangle+2\,{\rm Re}\,\langle 0|\chi_n^{(2)}\rangle=0\,,
\label{constr}
\end{align}
which follow from the normalization condition of $|\chi_n\rangle$ to the 1st-order in $\Delta t$. Henceforth, subscript $n$ will be dropped in $\ket{0_n}$, $|\chi_n^{(1)}\rangle$ and $|\chi_n^{(2)}\rangle$.

Plugging \eqref{Un-3} and \eqref{chi-exp} into \eqref{Kraus} and grouping together terms of the same order in $\sqrt{\Delta t}$, to leading order we get
\begin{align}
\hat K_k=\langle k|0\rangle+\hat K_k^{(1)}\sqrt{\Delta t}+\hat K_k^{(2)}\Delta t\,,\label{Kk-exp}
\end{align}
where
\begin{align}
\hat K_k^{(1)}=&\,\langle k|\chi^{(1)}\rangle-i  \sqrt{\gamma}\langle k|1\rangle\A  \,,\label{K1}\\
\hat K_k^{(2)}=&\,\langle k|\chi^{(2)}\rangle-i \left(\sqrt{\gamma} (\,\langle k|{\hat b}_n|\chi^{(1)}\rangle \hat {\cal A}^\dag  + \langle k|{\hat b}^\dag_n|\chi^{(1)}\rangle \,\hat {\cal A} ) \right. \nonumber\\
&\left.\,\,\,\,\,\,\,\,+\langle k|0\rangle ( \hat H_{\rm vac}-i\tfrac{\gamma}{2}\,\hat {\cal A}^\dag \hat {\cal A} )\!\right)\label{K2},
\end{align}
with time-bin operators ${\hat c}_\mu$ and emitter operators $\hat {\cal C}_\mu$ ($\mu,\mu'=1,2$) given by \eqref{cC}.
Replacing \eqref{Kk-exp} into the conditional map \eqref{proj} yields (to leading order)
\begin{align}
\tilde\rho_n=\,&|\langle 0|k\rangle|^2\,\rho_{n-1}+\left(\langle 0|k\rangle \,\hat K_k^{(2)}\rho_{n-1}+{\rm H.c.}\right)\Delta t\nonumber\\
&+\hat K_k^{(1)}\rho_{n-1}\hat K_k^{(1)^\dag}\!\Delta t.\label{tilderho}
\end{align}
Summing the right-hand side over $k$ (see Appendix \ref{app-jumps}, we end up with the Lindblad master equation \eqref{me-jumps-sum} 
[recall definition \eqref{DJ}], where the effective Hamiltonian $\hat H_{\rm eff}$ and jump operators $\hat J_k$ are respectively given by \eqref{Heff-sum} and \eqref{Jk-sum}
(note that $\hat J_k=\hat K_{k}^{(1)}$).
This equation is equivalent to the (white-noise, Gaussian) master equation \eqref{g-ME} [or \eqref{g-ME-rl} for $\gamma'=0$], but at variance with this is not expressed in terms of field moments (requiring instead a more detailed knowledge of the time-bin state). Thus, $\hat H_{\rm eff}$ and $\hat J_k$ in fact define an unraveling of the master equation corresponding to a desired photodetection scheme. Two comments follow. 
\\	
\indent First, there are white-noise Gaussian field states, such as squeezed and thermal states, for which the 0th-order term of expansion \eqref{chi-exp} differs from $\ket{0_n}$ since their infinite bandwidth corresponds to an infinite photon flux (see also \rref\cite{grossQubit2018}). \eqs\eqref{me-jumps-sum}, \eqref{Heff-sum} and \eqref{Jk-sum}  thus do not apply to such states.
\\
\indent  Second, for most physically relevant field states, in a single time bin the single-photon and two-photon amplitudes are at most of order $\sqrt{\Delta t}$ and $\Delta t$, respectively.
This means that only the outcomes $k=0$ and $k=1$ occur with meaningful probability, corresponding respectively to ``click'' and ``no-click'' outcomes (in line with most treatments of photodetection which indeed limit themselves to ``click''/``no-click'' outcomes at each elementary time interval $\Delta t$). Assuming a detector with perfect efficiency, the probability of a click in the time window $[t_{n-1},t_n]$ can be worked out from \eq\eqref{pk} for $k=1$ with the help of \eqref{Kk-exp} and retaining only terms up to first order in $\Delta t$ (see Appendix \ref{AppD}). This yields the detection probability rate
\begin{align}\label{pk1}
\frac{p_1}{\Delta t}= &
|\langle \chi^{(1)} | 1 \rangle|^2 
+
(i  \sqrt{\gamma} \langle 1|\chi^{(1)}\rangle \langle \A^\dag \rangle{+}{\rm c.c.})+  \gamma \langle \A^\dag \A \rangle\,.
\end{align}
When the field is in the vacuum state, this reduces to ${p_1}/{\Delta t}=\gamma \langle \A^\dag \A \rangle$.

As an illustration, in the next subsection we will show how the above applies to photon counting in the case of a coherent-state wavepacket.
\\
\\
\indent In the most general case of a mixed time-bin initial state, whose spectral decomposition reads $\eta_n=\sum_{\kappa} q_{\kappa} \ket{\chi_{\kappa}}_n\!\bra{\chi_{\kappa}}$ (with probabilities $q_\kappa$ fulfilling $\sum_\kappa q_\kappa=1$), the Kraus operators will be indexed not only by the measurement outcome but also by the eigenstate $\ket{\chi_\kappa}$,
\begin{equation}
\hat K_{k,\kappa}=\sqrt{q_\kappa}\bra{k}\hat U\ket{\chi_\kappa}\,,
\end{equation}
and \eqref{proj} turns into a sum over $\kappa$,
\begin{equation}
  \tilde\rho_n=\sum_{\kappa}\hat K_{k,\kappa} \,\rho_{n-1} \,\hat K_{k,\kappa}^\dag\,.\label{proj-mixed}
\end{equation}

\subsection{Photon counting for a coherent-state wavepacket}

\label{sec_Bar}
In the case of photon counting, the time-bin measurement basis are the Fock states $\{|k_n\rangle \}$ with $k=0,1,...\,$. As anticipated in Section \ref{sec-jump-summ}, a coherent-state wavepacket in terms of time modes \eqref{tm} reads \cite{LoudonQTL03}
\begin{equation}
|\xi\rangle=e\,^{\int \!{d}t\,\left(\xi_t \hat b^\dag_t-\xi_t^*  \hat b_t\right)}\,|0\rangle\,\label{xi}
\end{equation}
with $\xi_t$ the wavepacket amplitude in the time domain \cite{note-xi}. One can decompose the time integrals into a sum over intervals $\{[t_{n-1},t_n]\}$ and, if $\Delta t$ is small enough, in each interval replace $\xi_t\rightarrow \xi_n=\tfrac{1}{\Delta t}\int_{t_{n-1}}^{t_n}\!dt\,\xi_t$. This yields 
\begin{equation}
|\xi_n\rangle\simeq  e^{\sum_n( \xi_n\sqrt{\Delta t}\,\hat b_n^\dag-\xi_n^*\sqrt{\Delta t}\,  \hat b_n)}|0_n\rangle\,,
\end{equation}
hence \eqref{uncorr} holds for $\eta_n=\ket{\chi_n}\!\bra{\chi_n}$. Here, each $\ket{\chi_n}$ [\cf\eq\eqref{chin-coh-sum}] is a single-mode coherent state of the $n$th time bin with amplitude $\xi_n\sqrt{\Delta t}$.
Expanding $|\xi_n\rangle$ in powers of $\sqrt{\Delta t}$, one ends up with the effective Hamiltonian and jump operator in \eq\eqref{HJ-coh}.
Note that, in addition to ${\hat H}_{\rm vac}$, $\hat H_{\rm eff}$ features the standard drive Hamiltonian, arising from the first-order term $\hat V_n$ in the collision unitary \eqref{Un-2} \cite{note-delta}.
Also, note the c-number shift in $\hat J_1$. Analogous shifted jump operators, physically due to the coherent superposition of light emitted from $S$ and the incoming beam, were previously derived (for normal emitters) via the input-output formalism (see, e.g, \rrefs\cite{WisemanMilburnBook, ZhangarXiv17,ManzoniNatComm17, CilluffoJSTAT19}).

Finally, the photocounting rate \eqref{pk1} in this case is given by
\begin{align}
\frac{d p_1}{d t}& =|\xi_t|^2 -2 \sqrt{\gamma} \, {\rm Im}\,( \xi_t \langle \A^\dag \rangle )+  \gamma \langle \A^\dag \A \rangle\,,
\end{align}
where we used $|\chi_n^{(1)}\rangle=\xi_n\ket{1_n}$ and passed to the continuous-time limit.

\subsection{Bidirectional field}\label{sec-jumps-bi}

If $\ket{\chi_n}$ ($\ket{\chi'_n}$) denotes the initial state of the right-going (left-going time bin) [recall Section \ref{CM-bi}], \eq\eqref{chi-exp} is generalized as
\begin{align}
\ket{\chi , \chi'} &= \ket{0, 0}+ (\ket{0 , \chi'^{(1)}} + \ket{\chi^{(1)} , 0})\sqrt{\D}\nonumber
\\&
+
(\ket{0 , \chi'^{(2)}} + \ket{\chi^{(2)}, 0}  + \ket{\chi^{(1)} , \chi'^{(1)}} )\,\D \label{chi_exp2w}
\end{align}
(we adopt the compact notation $\ket{a,  b'} = \ket{a} \otimes \ket{b'}$).
Plugging \eqref{Vn-rl} in \eq\eqref{Un-2}, the collision unitary to the lowest order reads [\cf\eq\eqref{Un-3}]
\begin{align}
\hat U_n\simeq &\,\openone  -i \left( \sqrt{\gamma}\A \,\hat b_n^\dag {+} \sqrt{\gamma'} \A' \,\hat {b'}_n^\dag {+}\rm H.c. \right)\! \sqrt{\Delta t} \nonumber\\
&-i\left(\hat H_{\rm vac}{-}\tfrac{i}{2}\left( \sqrt{\gamma}\A \,\hat b_n^\dag {+} \sqrt{\gamma'} \A' \,\hat {b'}_n^{\dag} {+}\rm H.c. \right)^2 \right)\!\Delta t\label{Un-2w}
\end{align}
with $\hat {H}_{\rm vac}$ given by \eq\eqref{Hvac-rl}. 

Finally, a procedure analogous to that leading to \eqs\eqref{Heff-sum}-\eqref{Jk-sum} yields \eqs\eqref{H_2w} and \eqref{J_2w} (see Appendix \ref{app-jumps} for details).

A photodetection event now corresponds to a measurement of both the right- and left-going time bins in a basis $\ket{k,k'}$ with $\{\ket{k}\}$ ($\{\ket{k'}\}$) an orthonormal basis of the right-going (left-going) time bin. The generalization of expansion \eqref{Kk-exp} can be worked out like in the unidirectional case; its explicit expression is reported in Appendix \ref{app-jumps}.

\section{Conclusions} \label{conclusions}

In this paper, we formulated the collision-model-based description of quantum optics dynamics in the presence of many quantum emitters, each able to interact with a generally chiral field at many coupling points. The collisional picture maps the field into a stream of discrete time-bin modes interacting with the emitters in a conveyor-belt-like fashion. In the regime of negligible time delays (usual in most experiments) the dynamics is effectively represented as a sequence of pairwise collisions each between a field time bin and all the emitters collectively. These at once undergo an internal dynamics ruled by an effective second-order Hamiltonian describing dipole-dipole interactions. This Hamiltonian origins from the fact that the traveling time bin reaches the system's coupling points in sequence, no matter how short the delays. As such, the effective Hamiltonian depends on the coupling points topology. 
We applied the collisional picture to derive a general Lindblad master equation of a set of (generally) giant emitters coupled to a chiral waveguide for an arbitrary white-noise Gaussian state of the field. This combines into a single equation and extends a variety of master equations used in quantum optics and waveguide QED. In addition, building on previous work \cite{grossQubit2018}, we worked out a general recipe that, for a given photodetection scheme, returns the effective Hamiltonian and jump operators generating the ensuing quantum trajectories.

For the sake of argument and in order to keep the number of parameters at a reasonable level, throughout we considered identical quantum emitters each featuring a single transition coupled to the field and no external drive. Extending the theory so as to relax these assumptions is straightforward.

It is natural to compare the collision-model picture with the longstanding input-output formalism (IOF) of quantum optics (and methodologies underpinned by the latter such as the SLH approach \cite{CombesAdvPhyX17}). As anticipated, these are both grounded on the same microscopic model, in particular the white-noise coupling assumption, and share the field representation in term of temporal modes. On the other hand, some significant differences stand out. One especially evident is that in the IOF time is a continuous variable, while in the collisional picture one in fact works with a discrete (coarse-grained) dynamics taking the continuous-time limit only at the end. 

Notably, while in the IOF one evolves the operators in a Heisenberg-picture-like fashion, the collisional framework essentially deals with evolutions of states in the spirit of the Schr{\"{o}}dinger picture. The role of input and output field operators in the IOF is played by the initial and final state of the time bins $\eta_n$ and $\eta'_{n}$, while the central equation of IOF that connects the output field to the input field and system operator is replaced by the pairwise unitary $\hat U_n$ describing each collision. The collision unitary concept makes the collisional picture particularly advantageous to carry out tasks such as deriving in a natural way CPT master equations, jump operators or effective decoherence-free Hamiltonians \cite{CCC}. 
Moreover, the joint emitters-field dynamics is in fact mapped into an effective quantum circuit, which can help quantum simulations and allows to potentially take advantage of already developed quantum information/computing techniques. In this respect, it was recently proved \cite{grossQubit2018} that all quantum optics master equations and photon detection schemes for a single (normal) emitter can be simulated through a collision model with time bins replaced by qubits. The present work in fact extends the same property to many giant emitters.

Finally, although in this paper the illustrations of the general theory mostly targeted the open dynamics of the emitters, we stress that the collisional picture captures the joint dynamics including the field. The framework is thus potentially as useful in problems such as multi-photon scattering from atoms \cite{JacobsPRA20} or non-equilibrium thermodynamics of quantum optics systems (or generally bosonic baths) \cite{Alexia2020}.

\section*{Acknowledgements}
We acknowledge support from MIUR through project PRIN Project 2017SRN-BRK QUSHIP, the Canada First Research Excellence Fund and NSERC.
AC acknowledges support from the Government of the Russian Federation through Agreement No. 074- 02-2018-330 (2).
\bibliography{WQED}

\begin{thebibliography}{80}
\bibitem [{not({\natexlab{a}})}]{note-xi}%
  \BibitemOpen
  \href@noop {} {}\bibinfo {howpublished} {Note that $\hat b_t$
  [\cf\eq\eqref{tm}] are defined in terms of $\omega$ measured from $\omega_0$.
  Thus, $\xi_t$ specifies the wavepacket in a frame rotating at frequency
  $\omega_0$.} ({\natexlab{a}})\BibitemShut {NoStop}%
\bibitem [{not({\natexlab{b}})}]{note1}%
  \BibitemOpen
  \href@noop {} {}\bibinfo {howpublished} {In \eq\eqref{Ut2}, we neglected the
  contribution coming from interval between $[t/\Delta t]\Delta t$ and $t$
  since this vanishes in the limit $\Delta t\rightarrow 0$.}
  ({\natexlab{b}})\BibitemShut {NoStop}%
\bibitem [{not({\natexlab{c}})}]{note-sub}%
  \BibitemOpen
  \href@noop {} {}\bibinfo {howpublished} {Subscripts ``vac'',``th'' and ``sq''
  come from ``vacuum'', ``thermal'' and ``squeezed''. The corresponding
  Hamiltonian terms are indeed proportional to $[b_t,b^{\dag '}_t]$, $b^{\dag
  '}_t b_t$ and $b^\dag_t b^{\dag '}_t$, respectively, reflecting the fact that
  $\HH_{\rm vac}^{(1)}$ arises for any field state, while typical field states
  for which $\HH_{\rm th}^{(1)}$ ($\HH_{\rm sq}^{(1)}$) occurs are thermal
  (squeezed) states.} ({\natexlab{c}})\BibitemShut {NoStop}%
\bibitem [{not({\natexlab{e}})}]{note-delta}%
  \BibitemOpen
  \href@noop {} {}\bibinfo {howpublished} {Note that the $1/\sqrt{\Delta t}$
  factor in the coupling strength of $\hat V_n$ cancels with the $\sqrt{\Delta
  t}$ factor appearing in the amplitude of \eqref{chin-coh-sum}.}
  ({\natexlab{e}})\BibitemShut {NoStop}%
\end{thebibliography}

\appendix

\section{Derivation of the microscopic Hamiltonian}

For completeness, here we report the derivation of Hamiltonian \eqref{Htot}, with $\hat H_S$, $\hat H_f$ and $\hat V$ respectively given by \eqs\eqref{Hf}, \eqref{Hf-rl} and \eqref{V-rl}, through linearization of the field dispersion law (see also \eg \rref\cite{ShenPRA09I}).

Consider a one-dimensional bosonic field with normal-mode ladder operators $\hat a_k$ and $\hat a_k^\dag$ where $k$ is a (continuous) wavevector that can take both positive and negative values. Let $\omega_k$ with $\omega_k=\omega_{-k}$  be the dispersion law (for simplicity we consider time-reversal invariant fields, but a more general treatment is possible). 
The free field Hamiltonian can be written as
\begin{align}
\hat{H}_{f}= \int_{-\infty}^0 dk\,\, \omega_k \,\hat a_k^\dag \hat a_k+\int_{0}^\infty dk\, \,\omega_k \,\hat a_k^\dag \hat a_k\,. \label{Hf-2}
\end{align}
The field weakly and non-locally couples to $N_e$ quantum emitters, the $\ell$th coupling point of the $j$th emitter lying at position $x_{j\ell}$. The interaction Hamiltonian reads
\begin{equation}
\hat V=\sum_{j=1}^{N_e} \sum_{\ell=1}^{{\cal N}_j}\hat V_{j\ell}
\end{equation}
with
\begin{eqnarray}
\hat V_{j\ell}=&&\,\hat{A}_{j}^\dag\int_{0}^\infty dk\, \tfrac{g_k}{\sqrt{2\pi}}\,e^{i k x_{j\ell}}\hat{a}_k\nonumber\\
&&+\hat{A}_{j}^\dag\int_{-\infty}^0 dk\, \tfrac{g_k}{\sqrt{2\pi}}\,e^{i k x_{j\ell}}\hat{a}_k+ {\rm H.c.}\,,\,\, \label{V-2}
\end{eqnarray}
where $g_k$ is the coupling rate with mode $k$. The free Hamiltonian of the emitters $\hat H_S$ is given in \eq\eqref{Hf}. In \eqs\eqref{Hf-2} and \eqref{V-2}, we conveniently split each integral into a positive and a negative $k$'s contribution in a way that, once right- and left-going modes are introduced as $\hat b_k=\hat a_{k\ge 0}$ and ${\hat {{b}'}}_k=\hat a_{k< 0}$, the free-field and interaction Hamiltonian can be expressed as
\begin{eqnarray}
\hat{H}_{f}&=& \int_{-\infty}^0 dk\,\, \omega_k \,{\hat {{b}'}}_k^\dag {\hat {{b}'}}_k+\int_{0}^\infty dk\, \,\omega_k \,\hat b_k^\dag \hat b_k\,, \label{Hf-3}\\
\hat V_{j\ell}&=&\hat{A}_{j}^\dag\int_{0}^\infty dk\, \tfrac{g_k}{\sqrt{2\pi}}\,e^{i k x_{j\ell}}\hat{b}_k\nonumber\\
&&+\hat{A}_{j}^\dag\int_{-\infty}^0 dk\, \tfrac{g_k}{\sqrt{2\pi}}\,e^{i k x_{j\ell}}\hat{b}'_k+ {\rm H.c.}\,\,\, \label{V-3}
\end{eqnarray}
Since the coupling is weak, the emitters significantly interact only with a narrow field's bandwidth centered at the emitter frequency $\omega_0=\omega_{k_0}=\omega_{-k_0}$. Accordingly, the dispersion law and coupling rates are approximated as
\begin{align}
&\omega_{k\ge 0}\simeq \omega_0+v (k-k_0)\,,\,\,\,\omega_{k<0}\simeq \omega_0-v (k+k_0)\,,\\
&g_{k{\ge} 0}\simeq g_{k_0}=g\,,\,\,g_{k<0}\simeq g_{-k_0}=g'\,,
\end{align}
with $v=\partial_k \omega_{k}$ the field's group velocity. At the same time, the limits of integration in each integral in \eqs\eqref{Hf-3} and \eqref{V-3} can be extended to the entire real axis. Thereby, \eqref{Hf-3} and \eqref{V-3} are turned into
\begin{align}
\hat{H}_{f}= &\,\, \omega_0\int_{-\infty}^\infty dk \,(\hat b_k^\dag \hat b_k+{\hat {{b}'}}_k^\dag {\hat {{b}'}}_k)\nonumber\\
&+\int_{-\infty}^\infty dk\, v(k{-}k_0) \,\hat b_k^\dag \hat b_k- \int_{-\infty}^\infty dk\,v (k{+}k_0) \,{\hat {{b}'}}_k^\dag {\hat {{b}'}}_k\,, \\
\hat V_{j\ell}=&\hat{A}_{j}^\dag e^{i k_0 x_{j\ell}}\!\int_{-\infty}^\infty dk\, \tfrac{g}{\sqrt{2\pi}}\,e^{i (k-k_0) x_{j\ell}}\hat{b}_k\nonumber\\
&\,\,\,\,+\hat{A}_{j}^\dag e^{-i k_0 x_{j\ell}}\int_{-\infty}^\infty dk\, \tfrac{g'}{\sqrt{2\pi}}\,e^{i (k+k_0) x_{j\ell}}\hat{b}'_k+ {\rm H.c.}\,
\end{align}
Note the appearance of phase factors $e^{\pm i k_0 x_{j\ell}}$. Next, by making the variable change $k-k_0\rightarrow k$ in integrals featuring $\hat b_k$'s and $-(k+k_0)\rightarrow k$ in integrals featuring ${\hat {{b}'}}_k$'s, we get
\begin{align}
\hat{H}_{f}= &\,\,\omega_0\int_{-\infty}^\infty dk \,(\hat b_k^\dag \hat b_k+{\hat {{b}'}}_k^\dag {\hat {{b}'}}_k)\nonumber\\
&\,\,+\int_{-\infty}^\infty dk\, vk \,\hat b_k^\dag \hat b_k+ \int_{-\infty}^\infty dk\,v k \,{\hat {{b}'}}_k^\dag {\hat {{b}'}}_k\,, \\
\hat V_{j\ell}=\,\,&\hat{A}_{j}^\dag \,e^{i k_0 x_{j\ell}}\!\int_{-\infty}^\infty dk\, \tfrac{g}{\sqrt{2\pi}}\,e^{i k x_{j\ell}}\hat{b}_k\nonumber\\
&\,\,+\hat{A}_{j}^\dag \,e^{-i k_0 x_{j\ell}}\!\int_{-\infty}^\infty dk\, \tfrac{g'}{\sqrt{2\pi}}\,e^{-i k x_{j\ell}}\hat{b}'_k+ {\rm H.c.}\,\,, 
\end{align}
where we redefined the field operators as $\hat b_{k+k_0}\rightarrow \hat b_k$ and ${\hat {{b}'}}\!_{-(k+k_0)}\rightarrow {\hat {{b}'}}_k$. Finally, changing to the frequency domain $\omega$ we end up with
\begin{align}
\hat{H}_{f}&= \int_{-\infty}^\infty d\omega\, \omega \,\hat b_\omega^\dag \hat b_\omega+ \int_{-\infty}^\infty d\omega\, \omega \,{\hat {{b}'}}_\omega^\dag {\hat {{b}'}}_\omega\,, \\
\hat V_{j\ell}&=\hat{A}_{j}^\dag\,e^{i \omega_0 \tau_{j\ell}}\!\int_{-\infty}^\infty d\omega\, \sqrt{\tfrac{\gamma}{2\pi}}\,e^{i \omega \tau_{j\ell}}\hat{b}_\omega\nonumber\\
&\,\,\,\,+\hat{A}_{j}^\dag\,e^{-i \omega_0 \tau_{j\ell}}\!\int_{-\infty}^\infty d\omega\, \sqrt{\tfrac{\gamma'}{2\pi}}\,e^{-i \omega \tau_{j\ell}}\hat{b}'_\omega+ {\rm H.c.}\,\,, 
\end{align}
where $\tau_{j\ell}=x_{j\ell}/v$, $\hat b_\omega=\hat b_{k}/\sqrt{v}$, ${\hat {{b}'}}_\omega={\hat {{b}'}}_{k}/\sqrt{v}$, $\gamma=g^2/ v$ and $\gamma'=g'^2/ v$. Note the two equivalent ways to express the phase factors $e^{\pm i \omega_0 \tau_{j\ell}}=e^{\pm i k_0 x_{j\ell}}$.

\section{Terms $\HH^{(1)}_{\rm th}$ and $\HH^{(1)}_{\rm sq}$}
\label{app1}

As anticipated in the main text, for $\tau_{\nu}-\tau_{\nu-1}\ll \Delta t$ for all $\nu$'s (negligible time delays) and setting $\tau_{1}=0$, in \eqs\eqref{H1th} and \eqref{H1sq} all time delays can be neglected replacing $s-\tau_\nu$ ($s'-\tau_{\nu'}$) with $s$ ($s'$). This yields
\begin{eqnarray}
\HH^{(1)}_{\rm th}\simeq\,&& \tfrac{i \gamma}{2 \Delta t} \sum_{\nu \nu'} [\A_{\nu},\A^\dagger_{\nu'}] \! \int_{t_{n-1}}^{t_{n}}\! \!{d} s\! \int_{t_{n-1}}^{t_{n}}\!\! {d} s' {\rm sgn}(s'{-}s) \hat b_s^\dag \hat b_{s'} ,\,\,\,\,\,\,\,\,\label{Hth-app}\\
\HH^{(1)}_{\rm sq}\simeq \,&&\tfrac{i \gamma}{4 \Delta t} \sum_{\nu \nu'} [\A_{\nu},\A_{\nu'}] \! \int_{t_{n-1}}^{t_{n}}\! \!{d} s\! \int_{t_{n-1}}^{t_{n}}\!\! {d} s' {\rm sgn}(s'{-}s) \hat b^\dag_s \hat b^\dag_{s'} \nonumber\\
&&+\,{\rm H.c.}\,,
\end{eqnarray}
where we introduced the sign function to get more compact expressions. The integral in $\HH^{(1)}_{\rm sq}$ vanishes identically, due to the antisymmetry of the integrand under the exchange $s \rightleftarrows s'$. 
The same argument applies for a bidirectional field~[\cf\eqs\eqref{HHbisq}], in which case the integrals in $\HH^{(1)}_{\rm sq}$ features extra terms with the same symmetry.

To evaluate $\HH^{(1)}_{\rm th}$, we expand the field $\hat b_t$ in terms of time-bin modes $\hat b_{n,k}$ [\cf\eqs \eqref{fourier} and \eqref{rnk}]. This yields
\begin{align}
\hat H_{\rm th}=-\gamma\sum_{\nu,\nu'}[\A_{\nu},\A_{\nu'}^\dag]\sum_{k\neq 0}\frac{\hat{b}_{n,k}^\dag\hat{b}_{n,k}-\left(\hat{b}_{n,k}^\dag\hat{b}_{n,0}+{\rm H.c.}\right)}{2\pi k}\,.\label{Hth-bnk}
\end{align}
As discussed in Section \ref{sec-bnk}, for $\Delta t$ short enough, each mode $\hat b_{n,k\neq 0}$ is in its own vacuum state $\ket{0}_{n,k}$ [\cf\eq\eqref{field-state}]. Thus, effectively, $\hat H_{\rm th}=0$. 

For a bidirectional field, we will additionally expand $\hat b'_t$ in terms of left-going time-bin modes $\hat b'_{n,k}$ [defined in full analogy with \eqref{rnk}]. This results in an expression similar to \eqref{Hth-bnk}, featuring overall terms of type $\sim \hat{\beta}'^{\dag}_{n,k}\hat{\beta}_{n',k'}$ with $\beta, \beta'=b,b'$ and where at least one among $k$ and $k'$ is non-zero. Thus $\HH^{(1)}_{\rm th}$ is negligible when modes $b_{n,k\neq 0}$ and $b'_{n,k\neq 0}$ are in the vacuum state.

\section{Derivation of $\hat H_{\rm eff}$ and jump operators} \label{app-jumps}

Summing over $k$ the right-hand side of \eqref{tilderho} yields the CPT map ${\cal E}[\rho_{n-1}]=\sum_k \hat K_k\, \rho_{n-1}\hat K_k^\dag$. Using \eqref{Kk-exp}, this can be arranged as
\begin{align}
{\cal E}[\rho_{n-1}]=\,\rho_{n-1}&+\sum_{k}\,\left(\langle k|0\rangle \,\rho_{n-1}\, {\hat{K}_{k}^{(1)\dag}}{+}{\rm H.c.}\right)\!\sqrt{\Delta t}\nonumber\\
&+\sum_{k}\,\left(\langle k|0\rangle \,\rho_{n-1}\, {\hat{K}_{k}^{(2) \dag}}+{\rm H.c.}\right.\nonumber\\
&\left.\,\,\,\,\,\,\,\,\,\,\,\,\,\,\,\,\,\,+ \, \hat{K}_k^{(1)}\,\rho_{n-1}\,\hat{K}_k^{(1)\dag}\right)\Delta t\,.\label{Erho}
\end{align}
The contribution $\sum_{k}\left(\dots\right)\sqrt{\Delta t}$ vanishes since we can arrange the sum as
\begin{align}
\sum_{k}\left(\dots\right)=&2\,{\rm Re}\,\langle 0|\chi_n^{(1)}\rangle\,\rho_{n-1}\nonumber\\
&- i \left[\rho_{n-1}, \sqrt{\gamma}\left(\langle 0|\hat b_n|0\rangle\A^\dag+{\rm H.c.}\right)\right]\,.
\end{align}
This is zero due to \eq\eqref{constr} and, of course, $\langle 0|\hat b_n|0\rangle=0$.

The remaining terms in \eqref{Erho}, using $\Delta\rho_n={\cal E}[\rho_{n-1}]-\rho_{n-1}$, yield
\begin{align}
\frac{\Delta\rho_n}{\Delta t}=\rho_{n-1}\,\hat A^\dag+\hat A\,\rho_{n-1}+\sum_{k}  \hat J_{k}\,\rho_{n-1}\,\hat J_{k}^\dag\label{quasi}\,,
\end{align}
where we defined
\begin{equation}
\hat A=\sum_{k}\, \langle0|k\rangle \,\hat K_{k}^{(2)}\,,\,\,\,\hat J_k=\hat K_{k}^{(1)}\,\,.\label{Aop}
\end{equation}
We can set $\hat A=\hat R- i \hat H_{\rm eff}$ with $\hat R=\tfrac{1}{2}(\hat A+\hat A^\dag)$ and $\hat H_{\rm eff}=\tfrac{i}{2}(\hat A-\hat A^\dag)$. With this replacement, \eq\eqref{quasi} becomes
\begin{equation}
\frac{\Delta \rho_n}{\Delta t}=- i \,[\hat H_{\rm eff},\rho_{n-1}]+\left[\hat R, \rho_{n-1}\right]_+\!+\sum_{k}  \hat J_k\,\rho_{n-1}\,\hat J_k^\dag\label{lind}\,.
\end{equation}
Since map ${\cal E}$ is in particular trace preserving, ${\rm Tr}_S(\Delta \rho_n)$ always vanishes, that is
\begin{equation}
{\rm Tr}_S\left(\!\left[\hat R, \rho_{n-1}\right]_+\!{+}\sum_{k}  \hat J_k\,\rho_{n-1}\,\hat J_k^\dag\right)=0\,.
\end{equation}
Since this must hold for any $\rho_{n-1}$, the argument of the trace is zero, yielding $\hat R =-\tfrac{1}{2}\sum_{k} \hat J_k^\dag\hat J_k$ \cite{breuerTheory2007}. Replacing back in \eqref{lind}, we thus obtain the dissipator of \eq\eqref{me-jumps-sum}.
To work out $\hat H_{\rm eff}$, we explicitly calculate $\hat A$ [\cf\eq\eqref{Aop}] with the help of \eqref{cC} and \eqref{K2}, obtaining
\begin{align}
\hat A=&\langle0|\chi^{(2)}\rangle-i \sqrt{\gamma} \,\langle 1|\chi^{(1)}\rangle \,\hat {\cal A}^\dag -i \hat H_{\rm vac} -\tfrac{\gamma}{2} \,\hat {\cal A}^\dag \hat {\cal A} \,
\end{align}
(note that the last term is Hermitian).
Plugging this in $\hat H_{\rm eff}=\tfrac{i}{2}(\hat A-\hat A^\dag)$, neglecting an irrelevant constant term ${\rm Im}\,\langle0|\chi^{(2)}\rangle$, we end up with the effective Hamiltonian in \eq\eqref{Heff-sum}, which concludes the proof.
\bigskip

\subsection{Bidirectional field}

Generalizing \eqref{Kraus} as $\hat {K}_{k,k'}=\bra{k,k'}\!\hat{ U}_n\ket{\chi,\chi'}$, the low-order expansion of each Kraus operators reads [\cf\eq\eqref{Kk-exp}]
\begin{align}
\hat K_{k,k'}=\hat{K}_{k, k'}^{(0)}+\hat{K}_{k, k'}^{(1)}\,\sqrt{\Delta t}+\hat{K}_{k, k'}^{(2)}\,\Delta t\,.
\end{align}
Using \eqs\eqref{chi_exp2w}-\eqref{Un-2w} and defining left-going primed operators in full analogy with \eq\eqref{cC}, we get
\begin{align}
\hat{K}_{k, k'}^{(0)}=& \la k | 0 \ra \la k'  | 0  \ra\,,\\
\hat{K}_{k, k'}^{(1)}=& \la k | 0  \ra \la k' | \chi'^{(1)} \ra {+} \la k | \chi^{(1)} \ra \la k' | 0 \ra  \nonumber\\&
-i \left(\sqrt{\gamma} ~\!\la k | 1 \ra \la k' | 0\ra \A+ \sqrt{\gamma'} ~ \!\la k | 0 \ra  \la k' | 1\ra   \Ap \right)\,, 
\end{align}
\begin{widetext}
	\begin{align}
	\hat{K}_{k, k'}^{(2)} =\,& \la k | 0 \ra \la k' | \chi'^{(2)} \ra { +} \la k' | 0 \ra \la k | \chi^{(2)} \ra {+} \la k | \chi^{(1)} \ra \la k' | \chi'^{(1)} \ra 
	-i \bigg[
	\la k | 0 \ra \la k' | 0 \ra  \hat{H}_{\rm vac} -\tfrac{i}{2}  
	\left(\gamma\, \la k' | 0 \ra\! \la k|0 \ra \hat{\cal{A}}^\dag \hat{\cal{A}}  \nonumber
	\right.\\
	&\left.+ \gamma' \, \la k | 0 \ra \la k'|0 \ra   \Ap^\dag \Ap +2\sqrt{ \gamma \gamma'} \,  \la k' | 1 \ra  \la k | 1 \ra \Ap \A \right)\bigg]  
	- i \bigg [\sqrt{\gamma} ~ 
	\left((\la k | 1\ra \la k' | \chi'^{(1)} \ra  + \la k' | 0 \ra \la k| \hat{b}^\dag_n |\chi^{(1)} \ra )  \A \right.
	\nonumber
	\\&\left.
	+ \la k' | 0 \ra \bra{k} \hat{b}_n \ket{\chi^{(1)}} \A^\dag \right)
	+
	\sqrt{\gamma'} ~\left( (
	\la k'| 1 \ra \la k | \chi^{(1)} \ra
	+
	\la k | 0 \ra  \bra{k'} \hat{b}'^{\dag}_n \ket{\chi'^{(1)}}) \Ap
	+
	\la k | 0 \ra   \bra{k'} \hat{b}'_n \ket{\chi'^{(1)}} \Ap^\dag \right)\bigg]\,.
	\end{align}
\end{widetext}
Plugging these into \eq\eqref{Erho} with the replacements $\ket{0}\rightarrow\ket{0,0 }$, $k\rightarrow k,k'$ yields the CPT map at each collision. Terms $\sim \sqrt{\D}$ vanish using an argument analogous to the unidirectional case. 
Essentially for the same reason, when summing over $(k,k')$, one finds that crossed terms $\sim \Ap \A$ in $\hat{K}^{(2)}$ yield a zero contribution.
Finally, repeating a reasoning analogous to the unidirectional case lead to \eqs\eqref{H_2w} and \eqref{J_2w}.

\section{Derivation of \eq\eqref{pk1}}\label{AppD}

From \eq\eqref{Kk-exp} we get to leading order
\begin{align}
\hat{K}_k^\dag \hat{K}_k =&
 |\langle 0 | k \rangle|^2 + (\langle 0 | k \rangle \hat{K}^{(1)}_k + \langle k | 0 \rangle \hat{K}^{\dag(1)}_k)\sqrt{\Delta t}  \nonumber\\
&+
(\langle 0 | k \rangle \hat{K}^{(2)}_k 
+
\langle k | 0 \rangle \hat{K}^{\dag(2)}
+ 
\hat{K}^{\dag(1)}_k \hat{K}^{(1)}_k
) \Delta t \,.
\end{align}
For $k=1$, this reduces to $\hat{K}^{\dag(1)}_k \hat{K}^{(1)}_k \Delta t$. Applying \eq\eqref{K1} and dividing by $\Delta t$, we end uo with \eq\eqref{pk1}.
Analogously for bidirectional field, since $\hat{K}_{1, 0}^{(0)} = \hat{K}_{0, 1}^{(0)} = 0$, the only contributing term is $\hat{K}_{0, 1}^{\dag(1)} \hat{K}_{0, 1}^{(1)} 
+
\hat{K}_{1, 0}^{\dag(1)} \hat{K}_{1, 0}^{(1)} 
$, then
\begin{align}\label{pkk1}
\frac{p_1}{\Delta t}= &
|\langle \chi^{(1)} | 1 \rangle|^2 
+
(i  \sqrt{\gamma} \langle 1|\chi^{(1)}\rangle \langle \A^\dag \rangle{+}{\rm c.c.})+  \gamma \langle \A^\dag \A \rangle \nonumber
\\&
|\langle \chi'^{(1)} | 1 \rangle|^2 
+
(i  \sqrt{\gamma'} \langle 1|\chi'^{(1)}\rangle \langle \Ap^\dag \rangle{+}{\rm c.c.})+  \gamma' \langle \Ap^\dag \Ap \rangle
\,.
\end{align}

\end{document}